\documentclass[fleqn,usenatbib,letterpaper]{mnrasletter}

\usepackage[totalwidth=480pt,totalheight=680pt,layoutvoffset=0.5cm]{geometry}
\usepackage{newtxtext,newtxmath}

\usepackage[T1]{fontenc}
\usepackage{ae,aecompl}

\usepackage{graphicx}	
\usepackage{amsmath}	
\usepackage{amssymb}	
\usepackage{breqn}
\usepackage{multirow}

\usepackage{rotating}\usepackage{adjustbox}

\title[Debris discs in binary systems]{A statistically significant lack of debris discs in medium separation binary systems}

\author[B. Yelverton et al.]{
Ben Yelverton$^{1}$\thanks{E-mail: bmy21@cam.ac.uk},
Grant M. Kennedy$^{2,3}$,
Kate Y. L. Su$^{4}$ and
Mark C. Wyatt$^{1}$
\\
$^{1}$Institute of Astronomy, University of Cambridge, Madingley Road, Cambridge CB3 0HA, UK\\
$^{2}$Department of Physics, University of Warwick, Gibbet Hill Road, Coventry CV4 7AL, UK\\
$^{3}$Centre for Exoplanets and Habitability, University of Warwick, Gibbet Hill Road, Coventry CV4 7AL, UK\\
$^{4}$Steward Observatory, University of Arizona, 933 N Cherry Avenue, Tucson, AZ 85721, USA
}

\date{Accepted XXX. Received YYY; in original form ZZZ}

\pubyear{2019}

\begin{document}
\label{firstpage}
\pagerange{\pageref{firstpage}--\pageref{lastpage}}
\maketitle

\begin{abstract}
We compile a sample of 341 binary and multiple star systems with the aim of searching for and characterising Kuiper belt-like debris discs. The sample is assembled by combining several smaller samples studied in previously published work with targets from two unpublished \textit{Herschel} surveys. We find that 38 systems show excess emission at 70 or 100~$\mu$m suggestive of a debris disc. While nine of the discs appear to be unstable to perturbations from their host binary based on a simple analysis of their inferred radii, we argue that the evidence for genuine instability is not strong, primarily because of uncertainty in the true disc radii, uncertainty in the boundaries of the unstable regions, and orbital projection effects. The binary separation distributions of the disc-bearing and disc-free systems are different at a confidence level of $99.4\%$, indicating that binary separation strongly influences the presence of detectable levels of debris. No discs are detected for separations between $\sim$25 and 135~au; this is likely a result of binaries whose separations are comparable with typical disc radii clearing out their primordial circumstellar or circumbinary material via dynamical perturbations. The disc detection rate is $19^{+5}_{-3}\%$ for binaries wider than 135~au, similar to published results for single stars. Only $8^{+2}_{-1}\%$ of systems with separations below 25~au host a detectable disc, which may suggest that planetesimal formation is inhibited in binaries closer than a few tens of au, similar to the conclusions of studies of known planet-hosting binaries.

\end{abstract}




\section{Introduction}
\label{sec:intro}

The planet formation process, characterised by the accumulation of micron-sized dust particles in the protoplanetary disc into increasingly large bodies, produces a population of planetesimals with sizes of order 1--1000~km (\citealt{Johansen14_Planetesimals}). The primordial dust is expected to be removed from the disc by Poynting-Robertson (PR) drag within a few Myr (\citealt{Wyatt15_FiveSteps}), leaving behind these larger bodies. Yet, a significant proportion of stars older than $\sim$10~Myr show excess flux at infrared wavelengths, which is generally interpreted as thermal emission from circumstellar dust which has been heated by the stellar radiation (e.g. \citealt{Aumann84_Vega}; \citealt{Oudmaijer92_IRExcess}; \citealt{Mannings98_IRExcess}). Rather than being primordial, the dust responsible for this emission is understood to be continuously replenished by destructive collisions between the planetesimals (\citealt{Backman93_VegaPhenom}; \citealt{Dominik03_VegaTheory}). 

These discs of planetesimals and the products of their collisions are known as \textit{debris discs}, and can be distinguished observationally from primordial dust discs by their relatively low fractional luminosities (e.g. \citealt{Hughes18_Review}). Importantly, an infrared excess corresponding to a debris disc in a given system acts as a signpost that the process of planet formation was able to proceed to a significant degree in that system. The proportion of stars with detected debris discs has been estimated at around 25\% for A and F type stars, and close to 15\% for G and K types (\citealt{Su06_AStarDiscs}; \citealt{Hillenbrand08_FEPS}; \citealt{Trilling08_DDSunLike}; \citealt{Eiroa13_DUNESDiscFrac}; \citealt{Sierchio14_MIPSPhot}; \citealt{Thureau14_HerschelA}; \citealt{Sibthorpe18_HerschelFGK}).

Binary and higher-order multiple stars make up a large fraction of all known star systems; for example, \citet{Raghavan10_Multiplicity} estimate that 44\% of FGK stars have at least one stellar companion (see also \citealt{Duquennoy91_MultiOrbEl}; \citealt{Eggleton08_MultiCat}; \citealt{Duchene13_MultiRev}). Planets do not appear to be uncommon in such systems, with \citet{Horch14_PlanetsInBinaries} concluding that $\sim$40--50\% of \textit{Kepler} planet hosts are binaries, comparable with the binary fraction of field stars. Similarly, \citet{Armstrong14_BinaryPlanets} used \textit{Kepler} data to infer that the occurrence rate of circumbinary planets with radii above 6$R_{\oplus}$ and periods less than 300~d is consistent with single star rates. However, the planet occurrence rate is strongly dependent on stellar separation. For example, \citet{Kraus16_RuinousCloseBinaries} estimated that binaries closer than $\sim$50~au are around a third as likely as wider systems to host a planet. Thus, it is important to consider the problem of how planetary systems form and evolve around binary systems. Knowledge of the properties of debris discs in binary systems, and how they depend on the stellar orbits, can contribute to our understanding of how often, and where, we expect planetesimals to be able to form therein, with implications for planets.

Several previous works have addressed this issue. \citet{Trilling07_BinaryDebris} observed 69 AFGK binary and multiple systems with the Multiband Imaging Photometer for \textit{Spitzer} (MIPS; \citealt{Rieke04_MIPS}), to search for excess emission at 24 and 70~$\mu$m. Considering as `excesses' measurements at least 2$\sigma$ above the expected stellar flux, they found a rather high overall disc detection fraction of around 40\%. Their results also suggested that binaries with orbital separations of 3--50 au were less likely to host a disc than those with closer or wider separations, with the explanation that such binaries have separations comparable to typical debris disc radii, such that planetesimals in those systems would tend to be in an unstable location. Nonetheless, three of their excesses implied the presence of dust in dynamically unstable regions (\citealt{Holman99_Stability}), and they proposed that these could be explained as dust migrating inwards from a stable region due to PR drag. However, this mechanism likely cannot provide a general explanation for detectable levels of apparently unstable dust, as dust grains in detectable debris discs are expected to be removed via collisional evolution and radiation pressure on time-scales faster than PR drag can act (\citealt{Wyatt05_InsignificanceOfPR}).

\citet{Rodriguez12_BinaryDebris} approached the problem differently, starting by gathering a sample of 112 (largely AFGK) stars thought to host discs based on previous work, and found that 28 of them belonged to a binary or multiple system. The distribution of separations of the binaries in their sample showed fewer systems in the 1--100~au range than would be expected for a randomly selected sample, again suggesting that `medium' separation binaries are less likely to host discs.

The most extensive study of debris discs in binary and multiple systems performed to date has been that of \citet{Rodriguez15_BinaryDebris}. They started with a volume limited sample of 449 AFGKM stars observed by \textit{Herschel} as part of the DEBRIS programme (\citealt{Matthews10_DEBRIS}), of which they identified 188 as belonging to a binary or multiple system. Combining 100 and 160~$\mu$m photometry from \textit{Herschel}'s Photodetector Array Camera and Spectrometer (PACS; \citealt{Poglitsch10_PACS}) with archival photometry at other mid to far infrared wavelengths, and using a 3$\sigma$ excess criterion, they found a low overall disc detection rate of around 11\% for the binaries and multiples versus 21\% for the single stars in their sample, in contrast to the results of \citet{Trilling07_BinaryDebris}. Superficially, their results suggested that binaries with periods of around $10^3$--$10^5$~d (corresponding to semimajor axes of around 1--50~au for Sun-like stars, which make up the majority of their sample) are less likely to host debris discs, similar to the conclusions of previous work. However, using a Kolmogorov-Smirnov (KS) test, they demonstrated that the period distributions of their disc-bearing and disc-free binaries and multiples were not, in fact, different in a statistically significant way (with a $p$-value of 0.09).

In this paper, we compile a large sample of binary and multiple systems which have photometric measurements in the far infrared, comprising systems drawn from several sources. Our sample consists of most of the systems analysed by \citet{Trilling07_BinaryDebris}, \citet{Rodriguez12_BinaryDebris} and \citet{Rodriguez15_BinaryDebris}, with the addition of systems from two unpublished \textit{Herschel} surveys: \texttt{OT2_gkennedy_2}, which targeted visual binaries with well known orbits, and \texttt{OT1_jdrake01_1}, which targeted close binaries. Considering overlap between the various sources, there are 341 unique systems in the full sample. We aim to use this sample to quantify the incidence and properties of debris discs in binary and multiple systems with higher statistical significance than has been possible in previous work. 

We expect such a result to be possible firstly because we simply have a larger sample, and secondly because of the inclusion of the \texttt{OT2_gkennedy_2} systems in particular. As they have well known orbits, their semimajor axes are in the 1--50~au range (for wider binaries, observations spanning a very long time period would be required to deduce their orbits precisely). Thus, these systems lie in the separation range of greatest interest, where previous work suggests that the separation distributions of disc-bearing and disc-free systems may differ. The inclusion of the \texttt{OT2_gkennedy_2} data also allows us to examine separately those systems whose orbits are well known. 

For our analysis, we gather archival photometry for each system to create spectral energy distributions (SEDs), then model these in search of 3$\sigma$ excesses in \textit{Spitzer} 70~$\mu$m and \textit{Herschel} 70 and 100~$\mu$m photometry characteristic of cool debris. Although broadly similar SED modelling of many of our systems has been performed in previous studies, analysing our sample as a whole rather than relying on the results of previous work removes any issues of inconsistency -- for example, \citet{Trilling07_BinaryDebris} used 2$\sigma$ as their excess threshold while \citet{Rodriguez15_BinaryDebris} used 3$\sigma$.

In section~\ref{sec:sample}, we describe in detail the make-up of our sample. Section~\ref{sec:SED_modelling} explains our methods for searching for and characterising debris discs within the sample using SEDs. The results of this analysis, with a particular emphasis on disc stability and detection statistics, are presented in section~\ref{sec:results}. In section~\ref{sec:discussion} we consider how our results relate to those of studies of protoplanetary discs and planets in binaries, then discuss what can be learnt about close binaries from the \texttt{OT1_jdrake01_1} data. We present our overall conclusions in section~\ref{sec:conclusions}.

\section{The sample}
\label{sec:sample}

This section explains how our sample of binary and multiple star systems was compiled, before briefly presenting the orbital and stellar properties of the sample as a whole. As the aim of this paper is to search for and characterise excesses in the far infrared, where debris disc emission typically peaks, we compile systems that have been observed at 70~$\mu$m by MIPS or at 70 and/or 100~$\mu$m by PACS.

First, we include all 34 systems targeted by the \textit{Herschel} survey \texttt{OT2_gkennedy_2}. All were observed at 70 and 160~$\mu$m by PACS. The bulk of these systems were drawn from the Sixth Catalog of Orbits of Visual Binary Stars (VB6; \citealt{Hartkopf01_VBCat}). This survey aimed to obtain observations of binaries whose orbits are well known, and so systems with highly graded orbital solutions in VB6 (grade 1, `definitive', or grade 2, `good') were chosen. The list of targets was narrowed by retaining only those stars with spectral types AFG and luminosity classes V or IV-V, and within a distance of 40~pc. Further exclusions were made of stars with non-main sequence positions on a Hertzsprung-Russell diagram, and one system lying in the galactic plane. In addition, we excluded systems with existing far infrared observations: several with PACS observations from other programmes, one (HD~26690) with a MIPS~70$\mu$m observation from \citet{Trilling07_BinaryDebris} suggesting a lack of dust, and two (HD~40183 and HD~155125) with constraints from the \textit{Infrared Astronomical Satellite} (\textit{IRAS}).

The resulting sample contains 27 systems. Of the remaining seven targets in the programme, three (HD~46273, HD~80671 and HD~127726) are those systems suggested by \citet{Trilling07_BinaryDebris} to harbour unstable dust, and were included with the aim of resolving the proposed discs. The final four (HD~31925, HD~95698, HD~173608 and HD~217792) are systems with dust temperatures which were poorly constrained based on the data of \citet{Trilling07_BinaryDebris}, and were included to allow us to better characterise their possible excesses with the addition of 160~$\mu$m photometry.

To these we add the 51 systems observed at 70 and 160~$\mu$m by PACS as part of the \textit{Herschel} survey \texttt{OT1_jdrake01_1}. This survey was motivated by the work of \citet{Matranga10_CloseBinaries}, which suggested that $\sim$20--30\% of close binaries host large amounts of very warm dust (with temperatures above $\sim$1000~K), and interpreted this dust as a possible product of collisions between planetary bodies that have been destabilised by long-term changes in the binaries' orbits. The \texttt{OT1_jdrake01_1} survey was designed to be complementary to that work: it aimed to search for cooler dust components around close binaries in order to build a more complete picture of their typical planetary systems (\citealt{Drake10_DestroyersProposal}). As this cooler dust is indicative of a more typical debris disc scenario in which ongoing collisions between planetesimals replenish the dust -- and to our knowledge there exists no published analysis of these data -- we include these systems in our sample.

We also include almost all of the binary and multiple systems studied by \citet{Trilling07_BinaryDebris}, \citet{Rodriguez12_BinaryDebris} and \citet{Rodriguez15_BinaryDebris}. The total numbers of binaries and multiples considered in each of those works are 69, 28 and 188 respectively. There is some overlap between the various samples: 15 systems appear in two of the previously mentioned studies, and two systems appear in all three. Additionally, nine systems from \texttt{OT2_gkennedy_2} and one from \texttt{OT1_jdrake01_1} (HD~118216) are in the \citet{Trilling07_BinaryDebris} sample. There is also the special case of HD~131976 (in \texttt{OT1_jdrake01_1}) and HD~131977 (in \citealt{Rodriguez15_BinaryDebris}): the latter is the primary star of the system GJ~570, while the former is the secondary component, itself a spectroscopic binary (see e.g. VB6). Since for most binary and multiple systems we do not have separate far infrared observations of each component, for consistency we discard the secondary, HD~131976, from our sample; for completeness, we note that its SED does not show an infrared excess. We exclude also six systems from \citet{Rodriguez12_BinaryDebris} -- HD~1051, HD~8538, HD~125473, HD~138749, HD~150378 and HD~169022 -- as these only appear to have been observed in the far infrared by \textit{IRAS}. Owing to the low spatial resolution of \textit{IRAS}, its photometry is prone to contamination from background sources, so that any apparent \textit{IRAS} excess without confirmation by MIPS or PACS would be somewhat less believable. A further two systems are excluded from \citet{Trilling07_BinaryDebris} -- HD~61497 and HD~111066 -- since they were observed only at 24~$\mu$m during the survey presented in that paper and have not since been followed up at longer wavelengths by either MIPS or PACS.

Finally, we add the following nine systems with highly graded VB6 orbits to the sample. HD~128620 was observed as part of the DEBRIS survey, and HD~48915 as part of \textit{Herschel} calibration; both would have been targeted by \texttt{OT2_gkennedy_2} if not for these existing observations\footnote{Since we ultimately chose to focus on MIPS and PACS data, we do \textit{not} include in our study the two systems which were excluded from \texttt{OT2_gkennedy_2} because of their existing \textit{IRAS} photometry.}. The systems HD~15285, HD~79969, HD~115953, HD~148653, HD~165341, HD~184467 and HD~196795 were observed by PACS as part of the DUNES survey (\citealt{Eiroa10_DUNES}). All are of spectral type K or M, and therefore, despite their well known orbits, would not have been included in \texttt{OT2_gkennedy_2} even if they had not been targeted by DUNES. Nonetheless, as we are not requiring that our whole sample be composed only of AFG stars, we include them in our analysis.

The resulting full sample contains 341 unique systems. The stars in the sample are largely primaries, but two of them are secondaries. The first is HD~137392, which was targeted by \texttt{OT2_gkennedy_2} instead of its primary HD~137391 because the BaBb orbit of that system, but \textit{not} the AB orbit, is highly graded in VB6. The second is HD~14082B, which was included in the sample of \citet{Rodriguez12_BinaryDebris} since it is known that the secondary hosts a detectable disc while the primary HD~14082A does not. We note that an alternative approach to constructing a sample of binaries would be to use \textit{Gaia} Data Release 2 (\citealt{Gaia18_DR2}) astrometry to make a list of pairs of stars with common proper motion. However, for this study we require stars which have been observed in the far infrared, which would not generally be the case for the proposed \textit{Gaia} sample. Conversely, for those systems which do not have well-characterised orbits -- which constitute the majority of the sample, as discussed below -- one might hope to use the \textit{Gaia} proper motions to confirm or refute their binary nature. We do not, however, attempt to do so, since the stars we will study are generally bright, with over half of the sample having a \textit{V}-band magnitude below 6, where \textit{Gaia} astrometry can be unreliable (\citealt{Gaia18_DR2Astrometry}).

As we are interested in how the presence of debris depends on the stellar orbits, for each system in our sample we record the stellar separations $a$ and, where available, orbital eccentricities $e$; this information is presented in Table~\ref{tab:binaryorbits}. Systems containing three or more stars have two or more separations. For the systems from \citet{Trilling07_BinaryDebris}, \citet{Rodriguez12_BinaryDebris} and \citet{Rodriguez15_BinaryDebris}, we mostly use the values given in those papers; their separations are not necessarily true semimajor axes, but in many cases projected separations. Of these systems, for those which appear in the VB6 catalogue with grade 1 or 2 we preferentially use the VB6 values -- the separations of these systems are true semimajor axes and can be considered well known. In some cases, the orbits given in the paper from which we took a system are incomplete; where possible, we supplement these with VB6 elements, even if they are poorly graded. For example, the triple system HD~196885 is listed in \citet{Trilling07_BinaryDebris} as having a separation of $190^{\prime\prime}$. This is for the orbit of the AB pair, and VB6 gives an additional grade 5 orbit for the AaAb pair with a separation of $0.64^{\prime\prime}$, which we record for completeness. In addition, HD~181296 is listed with only one separation (199~au) in \citet{Rodriguez12_BinaryDebris}. This is in fact a triple system, with the more distant companion being HD~181327, and as the system does not appear in VB6 we take the wider separation ($\sim2\times10^4$~au) from \citet{Tokovinin18_MSC}.

\begin{table*}\centering 
\begin{adjustbox}{max width=\textwidth}\begin{tabular}{llcccccccc}
\hline
Name    & Source      & Distance / pc & Component & Grade & VB6  $a / ^{\prime\prime}$ & VB6 $a$ / au  & VB6 $e$      & Source $a$ / au & Source $e$  \\ \hline
\multicolumn{10}{c}{$\cdots$} \\
HD 81858 & Kennedy     & 33.2   & AB       & 2 & 0.8599  & 28.52 & 0.5619 &            &         \\
HD 81997 & Rodriguez15 & 17.3   & AB, AB-C &   &         &       &        & 4.44, 1137 & 0.33, - \\
HD 82328 & Rodriguez15 & 13.5   & AB       &   &         &       &        & 55.3       &         \\
HD 82434 & Kennedy     & 18.8   & AB       & 1 & 0.81    & 15.24 & 0.431  &            &         \\
HD 82885 & Rodriguez15 & 11.4   & AB       & 5 & 3.84    & 43.66 & 0.88   & 43.6       & 0.88    \\
HD 83808 & Trilling07  & 40.0   & AaAb    & 2 & 0.00446 & 0.18  & 0      & 0.19       &         \\
HD 85091 & Drake       & 44.0   & AB       &   &         &       &        & 0.046      &         \\
HD 86146 & Drake       & 28.8   & AB       &   &         &       &        & 0.103      &         \\
HD 88215 & Trilling07  & 27.7   & AB       &   &         &       &        & 0.2        &         \\
HD 89125 & Rodriguez15 & 22.8   & AB       &   &         &       &        & 168.6      & \\
\multicolumn{10}{c}{$\cdots$} \\
\hline
\end{tabular}\end{adjustbox}
\caption{Information on the orbits of the binaries and multiple systems in our sample. The columns from left to right show the name of each system; its `source', i.e. the paper or programme from which we took the system (see text of section~\ref{sec:sample} for details); its distance, calculated as the reciprocal of the parallax; the names of the components that the given orbits refer to; the VB6 orbit grades, semimajor axes and eccentricities where available; and the separations and eccentricities given in the source where available. For systems from the \texttt{OT1_jdrake01_1} programme, the `source' separations are the values that we estimated from their periods, as detailed in section~\ref{sec:sample}. Here we display a ten-line excerpt chosen to show systems from a variety of sources; the full 341-line table is available online in machine-readable form.}\label{tab:binaryorbits}
\end{table*}

To convert the VB6 angular separations into physical separations, we mainly use parallaxes from \textit{Hipparcos} (\citealt{vanLeeuwen07_Hipparcos})\footnote{For systems not in the \textit{Hipparcos} catalogue, we mostly use parallaxes from \textit{Gaia} Data Release 2 (\citealt{Gaia18_DR2}). For the few systems which appear in neither catalogue, we source parallaxes from \citet{Harrington80_USNavalPlx}, \citet{Geyer88_Plx}, \citet{Altena95_CatPlx}, \citet{Costa05_SolarNeighborhood_Plx}, \citet{Henry06_SolarNeighborhood_Plx} and \citet{Torres10_NormalStars_Plx}.}. For the \texttt{OT2_gkennedy_2} systems -- other than those which were re-observations of the \citet{Trilling07_BinaryDebris} survey -- we use VB6 orbital elements, since by construction they are highly graded. For the \texttt{OT1_jdrake01_1} systems, we first find orbital periods, mostly from the Ninth Catalogue of Spectroscopic Binary Orbits (SB9; \citealt{Pourbaix04_SB9})\footnote{For the seven systems not listed in SB9, we source periods from the Multiple Star Catalog (\citealt{Tokovinin18_MSC}), \citet{Balona87_SpecPd}, \citet{Frasca06_SpecPd} and \citet{Kiraga12_SpecPd}.}. We estimate their luminosities $L$ from SED fitting (see section~\ref{sec:SED_modelling} for details), calculate approximate masses $M$ assuming $L\propto M^{3.5}$, then use these to estimate stellar separations from the periods.

\begin{figure}
	\centering
    \hspace{-0.5cm}
	\includegraphics[width=0.5\textwidth]{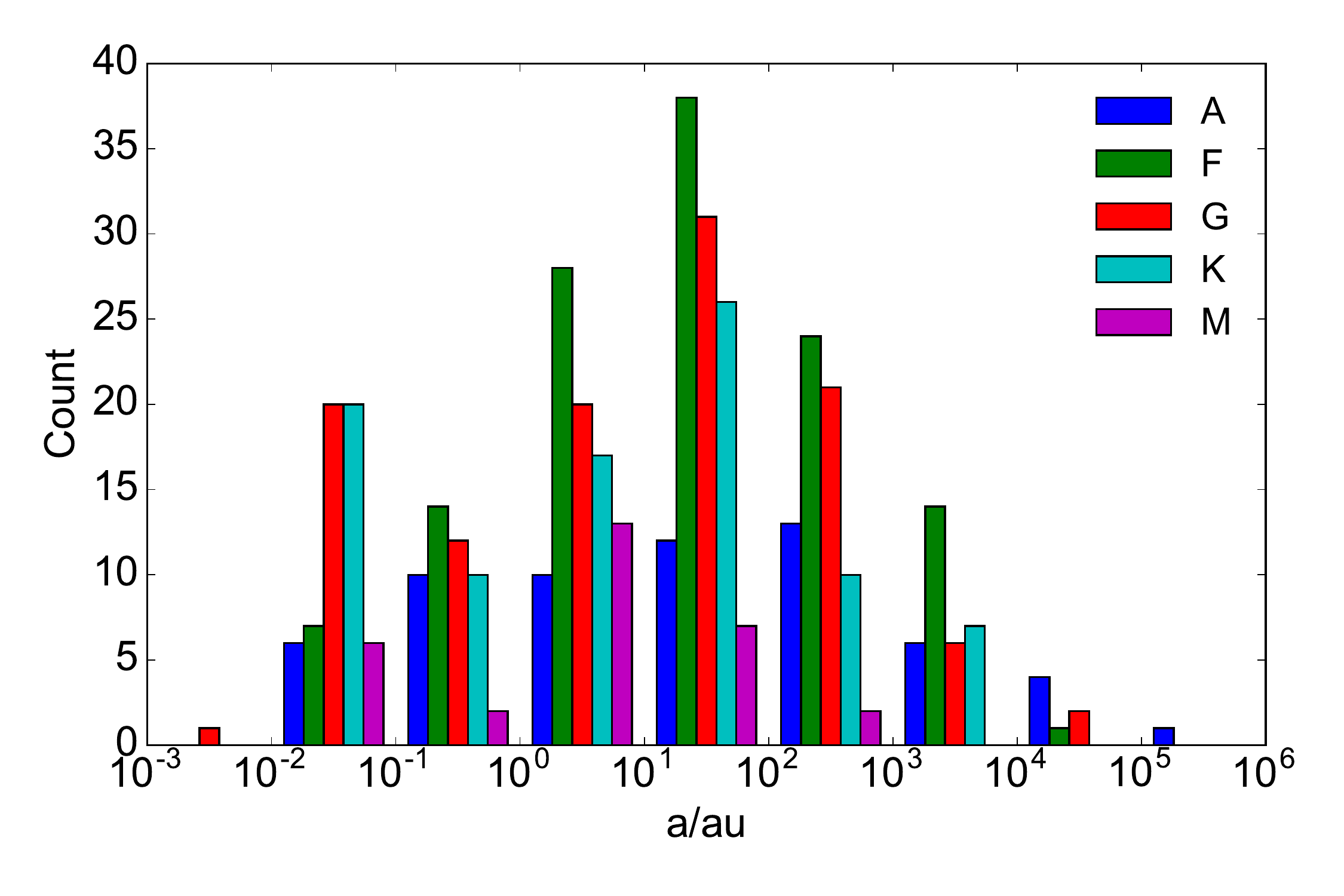}
	\caption{Distribution of stellar separations in our sample, binned by order of magnitude in separation and split by spectral type. Though there is some variation in the proportions of each spectral type between bins, the sample is largely Sun-like across the full range of separations.}
	\label{fig:sma_dist_per_sptype}
\end{figure}

The make-up of our sample is outlined in Fig.~\ref{fig:sma_dist_per_sptype}, which shows the distribution of stellar separations for each spectral type. The spectral types are derived by binning the stars by their effective temperature $T$ from SED fitting (again, see section~\ref{sec:SED_modelling} for details), with $T>7500~\mathrm{K}$ classified as type A, $6000~\mathrm{K}<T<7500~\mathrm{K}$ as F, $5200~\mathrm{K}<T<6000~\mathrm{K}$ as G, $3700~\mathrm{K}<T<5200~\mathrm{K}$ as K, and $T<3700~\mathrm{K}$ as M. Systems of three or more stars, and thus two or more separations, are counted multiple times. Note that the bimodality of the distribution is not representative of binary and multiple systems generally, but is largely the result of having included a significant number of binaries (from \texttt{OT1_jdrake01_1}) targeted specifically because they are close. 

Given that stellar ages are typically highly uncertain, in this paper we do not attempt to study disc properties as a function of age. As none of the samples from which we assembled our overall sample were selected based on age, in what follows we assume that stars of each spectral type have ages uniformly distributed up to the typical main-sequence lifetime of that type. Fig.~\ref{fig:sma_dist_per_sptype} shows that the sample is dominated by FGK stars, so we expect the average age of the systems we are studying to be a few Gyr.

To summarise, we have assembled a sample of 341 binary and higher order star systems and compiled information on their stellar separations; in most cases these are projected separations taken from the literature. We define a `known orbit' subsample consisting of all systems with at least one highly graded VB6 orbit and whose separations are therefore true semimajor axes; there are 89 such systems. For statistical analysis, care must be taken to avoid bias with respect to disc detection. Since \citet{Rodriguez12_BinaryDebris} studied only systems already thought to host discs, we will also define a `statistical' subsample consisting of all systems \textit{except} those taken from that paper. This subsample contains 328 systems (see section~\ref{sec:statistics}).  

Having established the systems to be studied in this paper, in the following section we outline our analysis of their SEDs and how we use these to determine whether a debris disc is present.

\section{SED Modelling}
\label{sec:SED_modelling}

Our initial aim is to establish whether or not each system in our sample hosts a detectable debris disc. This is equivalent to asking whether its SED indicates an excess of flux at infrared wavelengths, beyond what would be expected for a purely stellar source. To answer this, we first gather archival photometry for each system to create SEDs. The photometry available varies between systems, but all SEDs have a range of points spanning visible to far infrared wavelengths. By construction of the sample, all have a MIPS 70~$\mu$m, PACS 70~$\mu$m and/or PACS 100~$\mu$m observation; Table~\ref{tab:photometry} shows the photometry we use in these bands. For PACS photometry, we use our own flux measurements mostly from point spread function (PSF) fitting, and from aperture photometry for systems with resolved discs (such as HD~95698, as discussed in appendix~\ref{app:imagemodel}), as outlined in e.g. \citet{Sibthorpe18_HerschelFGK}. 

For MIPS, we use photometry from the literature in some cases, but for most systems we make our own updated flux measurements. The general procedure is outlined in \citet{Sierchio14_MIPSPhot}, where PSF photometry is used for unresolved sources. Since most of the MIPS observations were taken at both 24 and 70~$\mu$m, we generally use the MIPS 24~$\mu$m data to guide the position of the 70~$\mu$m detection. For sources not detected at 70~$\mu$m, we place an aperture at the 24~$\mu$m position to estimate the flux. As a result, such sources can have negative flux due to complex background structure and instrumental artefacts. Due to the large beam size of the MIPS 70~$\mu$m channel, in some cases only one MIPS 70~$\mu$m flux is extracted even if there are two MIPS 24~$\mu$m sources near the same position. 

\begin{table*}
\centering
\begin{tabular}{lcccccccccccccc}
\hline
  &\multicolumn{4}{c}{MIPS 70~$\mu$m}  && \multicolumn{4}{c}{PACS 70~$\mu$m} && \multicolumn{4}{c}{PACS 100~$\mu$m} \\
 
 \cline{2-5}  \cline{7-10}  \cline{12-15} \\

\multirow{2}{*}{Name} & $F_{\nu,\mathrm{obs}}$ & $\sigma_{\mathrm{obs}}$ & $F_{\nu,\mathrm{pred}}$ & \multirow{2}{*}{$\chi$} && $F_{\nu,\mathrm{obs}}$ & $\sigma_{\mathrm{obs}}$ & $F_{\nu,\mathrm{pred}}$ & \multirow{2}{*}{$\chi$} && $F_{\nu,\mathrm{obs}}$ & $\sigma_{\mathrm{obs}}$ & $F_{\nu,\mathrm{pred}}$ & \multirow{2}{*}{$\chi$} \\  

& / mJy & / mJy & / mJy &  && / mJy & / mJy & / mJy &  && / mJy & / mJy & / mJy &  \\ \hline

\multicolumn{15}{c}{$\cdots$} \\
HD 81858 &            &                  & 20.5           &        && 21.59      & 1.71             & 21.64          & -0.03  &&             &                   & 10.67           &         \\
HD 81997 & 33.84      & 3.58             & 31.79          & 0.57   &&            &                  & 33.56          &        && 16.16       & 2.65              & 16.46           & -0.11   \\
HD 82328 & 143.6      & 10.21            & 132.05         & 1.12   &&            &                  & 139.39         &        && 69.55       & 4.09              & 68.37           & 0.29    \\
HD 82434 & 78.06      & 5.32             & 70.14          & 1.43   && 79.07      & 2.75             & 74.06          & 1.55   &&             &                   & 36.72           &         \\
HD 82885 & -87.13     & 51.73            & 28.39          & -2.23  &&            &                  & 29.97          &        && 12.53       & 1.79              & 14.79           & -1.26   \\
HD 83808 & 90.24      & 5.41             & 90.2           & 0.01   &&            &                  & 95.22          &        &&             &                   & 46.99           &         \\
HD 85091 &            &                  & 3.55           &        && 3.82       & 2.17             & 3.75           & 0.03   &&             &                   & 1.85            &         \\
HD 86146 & 21.4       & 4.68             & 19.89          & 0.32   && 20.65      & 1.97             & 21             & -0.18  &&             &                   & 10.34           &         \\
HD 88215 & 22.25      & 3.63             & 13.13          & 2.5    &&            &                  & 13.86          &        &&             &                   & 6.81            &         \\
HD 89125 & 22.76      & 12.71            & 12.5           & 0.81   &&            &                  & 13.19          &        && 5.64        & 2.13              & 6.46            & -0.38  \\
\multicolumn{15}{c}{$\cdots$} \\
\hline
\end{tabular}
\caption{Photometry in the MIPS 70~$\mu$m, PACS 70~$\mu$m and PACS 100~$\mu$m bands for all systems in our sample. For each band, the columns show the observed flux density, its associated uncertainty, the predicted photospheric flux density, and the significance as defined in equation~(\ref{eqn:chidef}). For consistency, here we display the data for the same ten systems as in Table~\ref{tab:binaryorbits}; the full 341-line table is available online in machine-readable form.}\label{tab:photometry}
\end{table*}

To each SED, we then fit a model with both a stellar and a dust component, using the \textsc{MultiNest} algorithm \mbox{(\citealt{Feroz09_MultiNest})}\footnote{Our SED fitting software, \textsc{sdf}, is available at \url{https://github.com/drgmk/sdf}}. For the stellar component, we use \textsc{PHOENIX} BT-Settl models (\citealt{Allard12_PHOENIX}) with stellar abundances from \citet{Asplund09_Abundances}, as these are available over a wide range of parameters. Although some binaries will be sufficiently close that both stars contribute to the total measured flux, in all cases we fit only a single stellar component. We do not expect this simplification to affect our results significantly, because if the stars have different spectral types then the primary star will dominate the emission so that the secondary can be ignored. If both stars are of the same spectral type and thus both contribute significantly to the total flux, then the overall spectrum will be well approximated by a single star of the same spectral type. In addition, most systems have multiple mid infrared flux measurements, meaning that the Rayleigh-Jeans tail -- the part of the stellar spectrum that will be compared with the measured far infrared flux to search for an excess -- will be well constrained regardless of the details of the stellar model. We model the dust as a modified black body -- i.e. its emission is given by a black body function multiplied by a factor $(\lambda_0/\lambda)^\beta$ for wavelength $\lambda>\lambda_0$, with $\lambda_0$ and $\beta$ (as well as the dust temperature and luminosity) as free parameters. Physically, this extra factor accounts for the fact that the disc emission originates from a distribution of dust grain sizes, with grains emitting less efficiently at wavelengths larger than their own size. We model a single system, HD~95698, with two separate dust components (see e.g. \citealt{Kennedy14_TwoTemp}), as this gives a significantly better fit to its \textit{Spitzer} Infrared Spectrograph (IRS; \citealt{Houck04_IRS}) spectrum.

Our approach differs slightly from the most common method of SED fitting, which is to choose some cut-off wavelength ($\sim$10~$\mu$m), first using all photometry at shorter wavelengths to fit the stellar component, then extrapolate the resulting stellar model to far infrared wavelengths. The choice of cut-off wavelength could, however, affect the stellar model, for example if warm dust were present in a system, resulting in excess mid infrared flux. Fitting the stellar and dust components simultaneously allows us to obtain a far infrared photospheric flux prediction \textit{without} the need to choose a cut-off wavelength. Most of the `dust' components will in fact be insignificant -- that is, consistent with zero emission, and barely contributing to the total flux. However, for systems whose excesses we conclude \textit{are} significant, the dust model can then be used to characterise the inferred debris disc.

Thus, it must next be decided whether each SED should be considered to show a significant excess of infrared emission. We choose to focus only on excesses at 70 and 100~$\mu$m, corresponding to typical dust temperatures of order 30--200~K. It is possible for systems to host warmer dust, which would be better searched for at mid infrared wavelengths. However, such warm dust cannot generally be a direct tracer of planetesimal belts in the same way as the cooler dust that we are searching for. This is because a belt of planetesimals at the inferred distance of the warm dust should generally have collisionally depleted to below observable levels on a short time-scale (e.g. \citealt{Wyatt08_Review}). It has been suggested that warm dust could either have originated from planetesimals scattered inwards from a more distant belt (\citealt{Wyatt07_HotDust}) or in situ from collisions between planetary embryos (\citealt{Rhee08_WarmDust}). Systems with excesses in the mid infrared \textit{only} are rare (e.g. HD~69830: \citealt{Lisse07_HD69830}; \citealt{Marshall14_PlanetsAndDebris}). Note that BD+20~307, which is well known to host warm dust (\citealt{Song05_BD20307}), is in our sample, though this system also has an excess of far infrared emission. The key point is that warm dust and cool dust can be regarded as distinct phenomena, and so for consistency and ease of interpretation we restrict our attention to the latter. While we only search for excesses at 70 and 100~$\mu$m, shorter wavelength photometry -- for example from the Wide-field Infrared Survey Explorer (WISE; \citealt{Wright10_WISE}) 22~$\mu$m and MIPS 24~$\mu$m bands -- is useful for constraining dust temperatures where an excess is detected.

\begin{figure}
	\centering
    \hspace{-0.5cm}
	\includegraphics[width=0.5\textwidth]{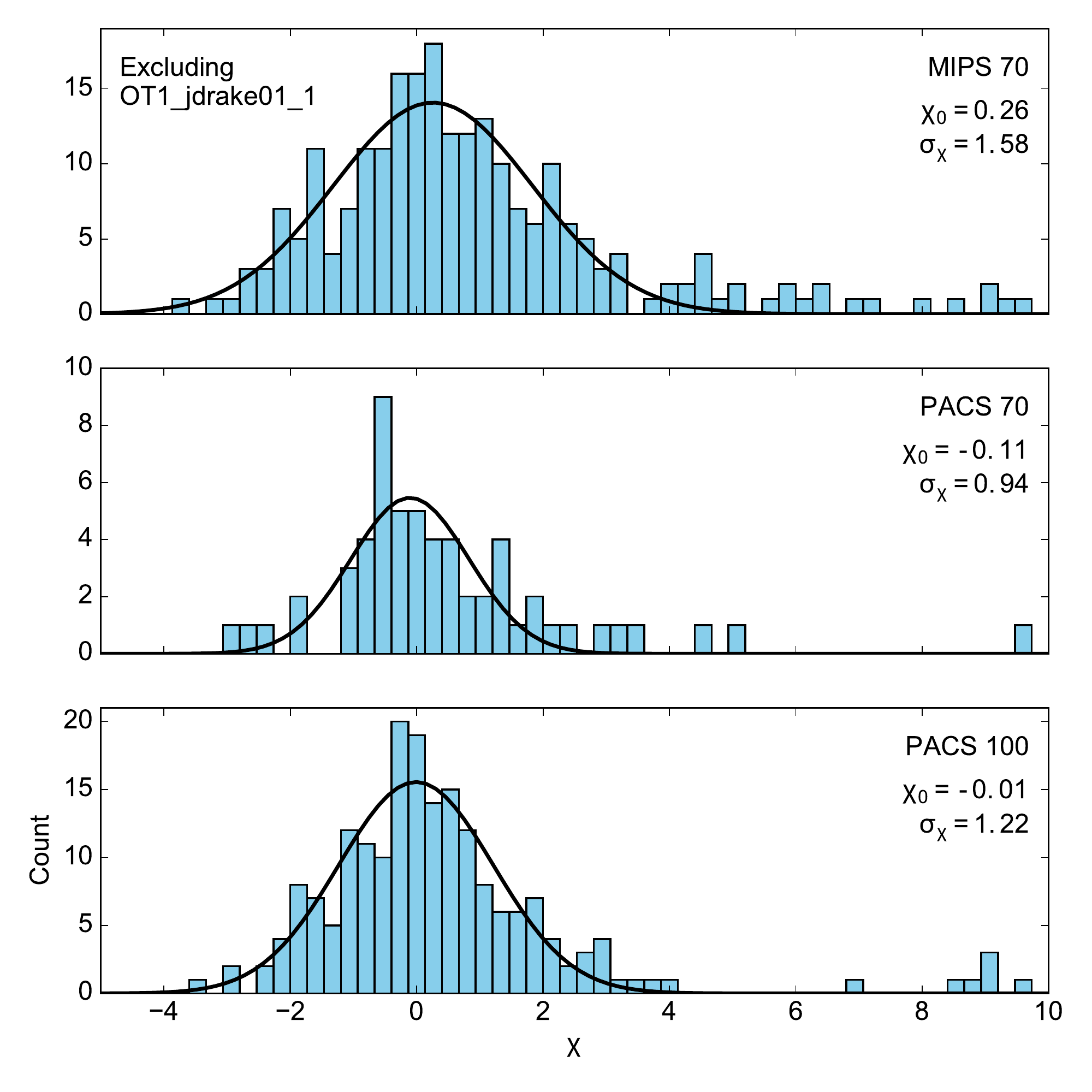}
	\caption{Distributions of photometric significance $\chi$ (bars) for all systems in our sample other than those from the \texttt{OT1_jdrake01_1} survey, with best fitting Gaussians (curves) overplotted. Values of $\chi$ greater than 10 are not shown. The upper, middle and lower panels show photometry from the MIPS 70~$\mu$m, PACS 70~$\mu$m and PACS 100~$\mu$m bands respectively. The annotations display the means $\chi_0$ and standard deviations $\sigma_\chi$ of the Gaussian fits.}
	\label{fig:chihist_nodrake}
\end{figure}

\begin{figure}
	\centering
    \hspace{-0.5cm}
	\includegraphics[width=0.5\textwidth]{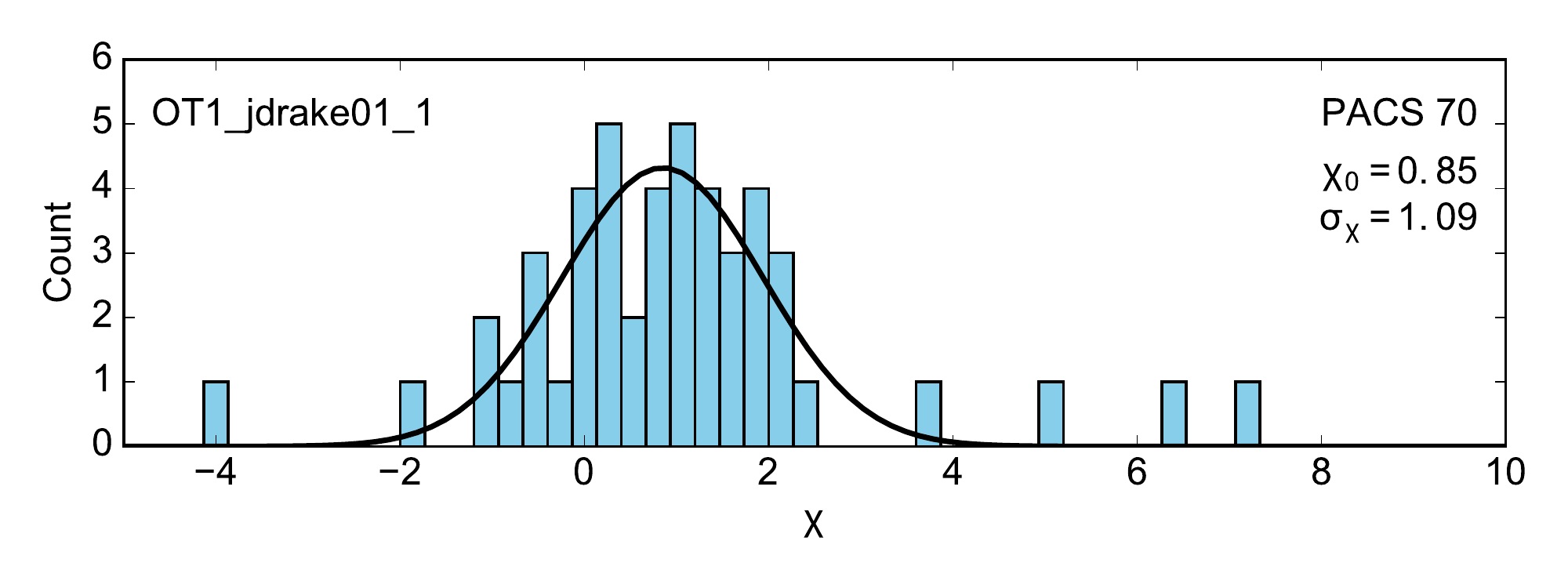}
	\caption{Same as Fig.~\ref{fig:chihist_nodrake}, but only for systems from the \texttt{OT1_jdrake01_1} survey. Only PACS 70~$\mu$m photometry is shown, as only two of these systems have MIPS 70~$\mu$m observations and none have PACS 100~$\mu$m observations.}
	\label{fig:chihist_drake}
\end{figure}

A useful measure of the significance of an excess at a given wavelength $\lambda$ is the quantity $\chi$, defined as:

\begin{equation}\label{eqn:chidef}
    \chi=\frac{F_{\nu,\mathrm{obs}}-F_{\nu,\mathrm{pred}}}{\sqrt{\sigma_{\mathrm{obs}}^2+\sigma_{\mathrm{pred}}^2}},
\end{equation}

where $F_{\nu,\mathrm{obs}}$ is the observed flux density at wavelength $\lambda$, $F_{\nu,\mathrm{pred}}$ is the flux density expected based on the best fitting stellar model alone, and $\sigma_{\mathrm{obs}}$ and $\sigma_{\mathrm{pred}}$ are the associated uncertainties. The values of $F_{\nu,\mathrm{pred}}$ and $\chi$ in the three far infrared bands we are focusing on are shown in Table~\ref{tab:photometry}, along with the measured photometry. For almost all systems, the $\sigma_{\mathrm{pred}}$ values (which are not shown) are around 1\% of $F_{\nu,\mathrm{pred}}$. For a given band, assuming no systematic offsets and accurately estimated uncertainties, the $\chi$ values for an unbiased set of observations would be expected to follow a near Gaussian distribution with mean $\chi_0=0$ and standard deviation $\sigma_\chi=1$, with any observations with $\chi>3$ considered to represent significant excesses. A relatively small proportion of stars host detectable levels of debris -- around 20\% for single stars, as discussed in section~\ref{sec:intro} -- hence the approximately Gaussian shape. In practice, we will use the observed $\chi$ distributions to define our excess criterion, as doing so automatically accounts for underestimated or overestimated uncertainties and systematic effects -- issues which may not be apparent when fitting individual SEDs.

In Fig.~\ref{fig:chihist_nodrake}, we show $\chi$ distributions for the three bands in which we will search for excesses, along with the best fitting Gaussians calculated using the least squares method. Note that this Figure excludes systems from the \texttt{OT1_jdrake01_1} survey; we find that the significances for those systems are distributed quite differently to the rest of the sample, and will therefore treat them separately (see below). Fig.~\ref{fig:chihist_nodrake} shows that for MIPS 70~$\mu$m, and to a lesser extent PACS 100~$\mu$m, the $\chi$ distributions for the bulk of the sample are somewhat broader than expected, with $\sigma_\chi>1$. This suggests that the uncertainties on either the predicted or observed fluxes in those bands have been underestimated. It could be the case that our estimates of the repeatability of the observations, particularly for MIPS 70~$\mu$m, are simply too optimistic. Some other possible explanations for the broadness of the distributions lie in the fact that our sample is comprised of binaries: firstly, the fact that we model each system with only a single stellar component could introduce extra uncertainty on the model fluxes. Secondly, some systems will have angular separations such that they are resolved into individual components at optical wavelengths, but not in the far infrared by MIPS and PACS due to the relatively large beam sizes of these instruments. Thus, some subset of systems will appear to have higher far infrared significances than they really do -- an effect which will tend to both broaden and positively offset the $\chi$ distributions. At 18$^{\prime\prime}$, the MIPS 70~$\mu$m beam size is the largest of the three far infrared bands, so this resolution effect should have the greatest impact on the MIPS 70~$\mu$m significance distribution, consistent with Fig.~\ref{fig:chihist_nodrake}.

In Fig.~\ref{fig:chihist_drake}, we plot the PACS 70~$\mu$m $\chi$ distribution for the \texttt{OT1_jdrake01_1} systems alone; there are insufficient data in the other two bands to make useful histograms for these systems. This distribution has a large positive offset $\chi_0=0.85$, which is not present in the rest of the sample (cf. Fig.~\ref{fig:chihist_nodrake}). \citet{Matranga10_CloseBinaries} observed that close binaries often have warm infrared excesses, which are not due to debris discs in the typical sense, and the offset we observe in the \texttt{OT1_jdrake01_1} PACS data could be related to that phenomenon (see section~\ref{sec:closebinaries}).

To account for the different spreads and offsets of the distributions, for the bulk of the sample, we require a system to have $\chi>\chi_0+3\sigma_\chi$ in at least one of MIPS 70~$\mu$m, PACS 70~$\mu$m and PACS 100~$\mu$m for its excess to be considered significant. We additionally require that ${\chi>3}$, since if this is not satisfied then the dust temperature of the model will be poorly constrained by the SED fitting. This second criterion is only relevant to PACS 70~$\mu$m, as the other two bands have $\chi_0+3\sigma_\chi>3$. In practice, requiring ${\chi>3}$ only eliminates a single system which would otherwise be considered to host a disc: HD~81809, which has $\chi=2.90$ in PACS 70~$\mu$m. For the \texttt{OT1_jdrake01_1} systems, we search only for PACS 70~$\mu$m excesses, using the same criteria but different distribution parameters as shown in Fig.~\ref{fig:chihist_drake}. Note that we treat the bands independently, without giving special treatment to systems with excesses in multiple bands. For example, while it could be argued that a system with significances below but close to the threshold in two different bands shows sufficient evidence for a disc, we would not classify such a system as disc-bearing using our criteria. As different systems in our sample have observations in different bands, taking such cases into account consistently would require defining different excess criteria for each possible combination of observations in the three far infrared bands.

There is one system in the sample to which we give special consideration in deciding whether a disc is present. HD~19356 (Algol) has a significant MIPS 70~$\mu$m excess with $\chi=5.12$, which could be linked to the system's variable millimetre emission rather than originating from a circumstellar disc (\citealt{Holland17_SONS}). Though there is no direct evidence against the presence of dust in the system, we take a conservative approach and, for the purpose of this paper, do not consider this system to host a debris disc.

In the following section, we present the results on the occurrence and properties of debris discs in our sample that follow from the SED analysis described above.

\section{Results}
\label{sec:results}

Using the methods outlined in the previous section, we are able to come to a conclusion about which systems in our sample are seen to host Kuiper belt-like debris discs, and to quantify the temperatures and luminosities of those discs. In this section, we first present these inferred disc properties and consider their implications for dynamical stability, then perform a statistical analysis to investigate whether the stellar orbits affect the detectability of debris in a significant way.

\subsection{Disc properties}
\label{sec:discproperties}

\begin{figure}
	\centering
	\includegraphics[width=0.5\textwidth]{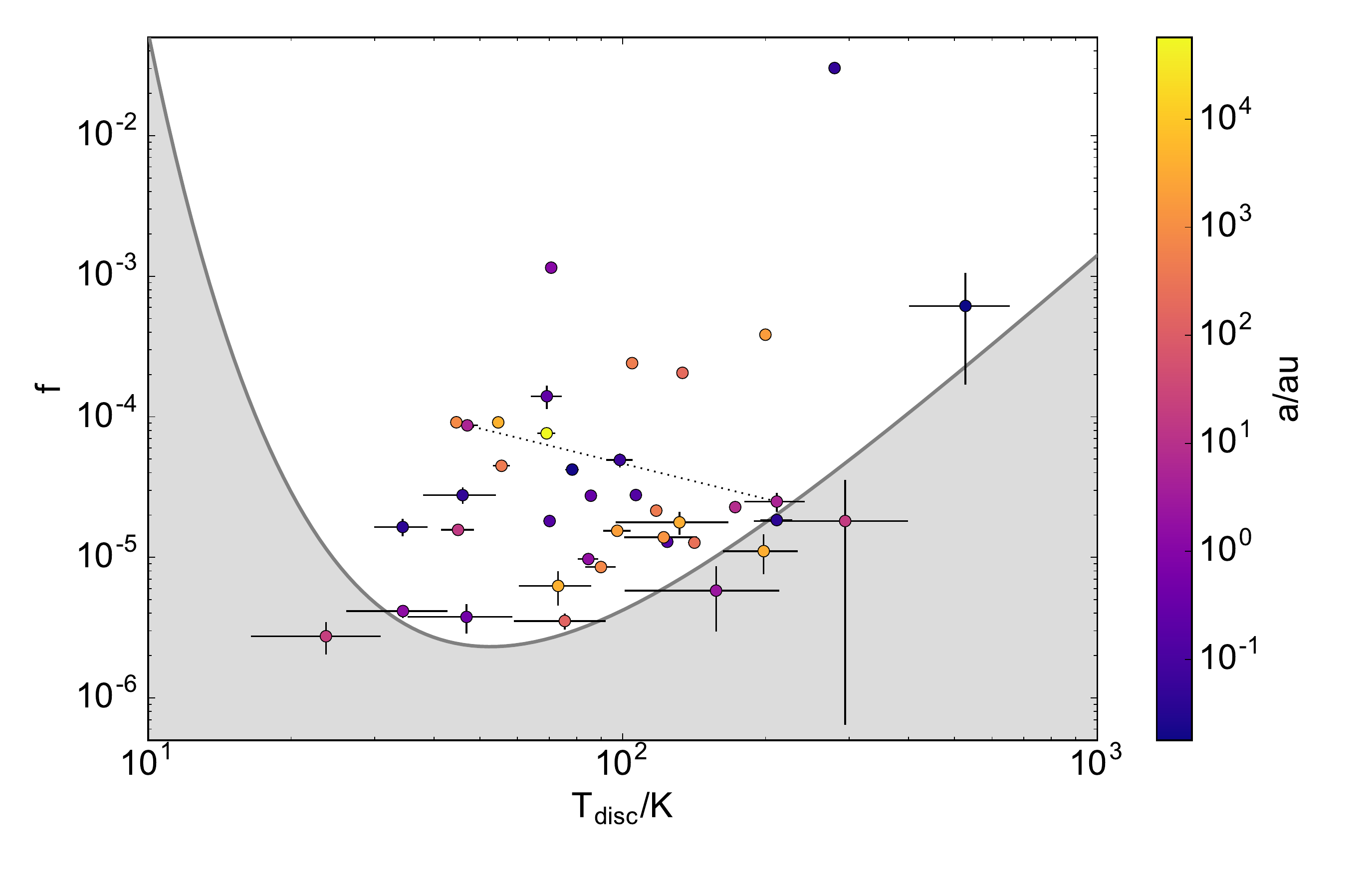}
	\caption{Fractional luminosity versus disc temperature for our full sample. The points are coloured by stellar separation; we use the smallest separation in the system for triple and higher order systems. We model HD~95698 as a two-temperature disc, and the two components are joined with a dotted line. The grey region shows the part of parameter space in which a disc around a typical star in our sample would be undetectable (see section~\ref{sec:discproperties} for details); systems lying close to the boundary of this region have comparatively large uncertainties.}
	\label{fig:fvsT}
\end{figure}

\begin{table*}
\begin{tabular}{lrrrrrrrrr}
\hline
Name      & $L/L_{\odot}$ & $T_{\mathrm{disc}}$/K  & $\Delta T_{\mathrm{disc}}$/K & $f$        & $\Delta f$  &$r_{\mathrm{bb}}$/au & $\Delta r_{\mathrm{bb}}$/au & $r_{\mathrm{disc}}$/au & $\Delta r_{\mathrm{disc}}$/au \\
\hline
HD 1404   & 23.1 & 142 & 4   & 1.3E-05 & 9.2E-07 & 19   & 1    & 34   & 2   \\
HD 8997   & 0.5  & 46  & 8   & 2.8E-05 & 3.7E-06 & 26   & 10   & 177  & 72  \\
HD 11171  & 5.8  & 73  & 13  & 6.3E-06 & 1.7E-06 & 36   & 14   & 105  & 42  \\
HD 13161  & 72.8 & 86  & 2   & 2.7E-05 & 6.7E-07 & 89   & 4    & 108  & 5   \\
HD 14082B & 1.1  & 105 & 2   & 2.4E-04 & 8.0E-06 & 7.3  & 0.2  & 39   & 1   \\
HD 16628  & 37.6 & 97  & 7   & 1.6E-05 & 1.4E-06 & 50   & 7    & 77   & 10  \\
HD 17094  & 10.8 & 157 & 56  & 5.8E-06 & 2.8E-06 & 10   & 8    & 23   & 19  \\
HD 17925  & 0.4  & 99  & 6   & 4.9E-05 & 5.7E-06 & 5    & 1    & 37   & 5   \\
HD 19994  & 4.1  & 34  & 8   & 4.2E-06 & 4.4E-07 & 146  & 66   & 484  & 220 \\
HD 20320  & 11.7 & 70  & 2   & 1.8E-05 & 8.9E-07 & 54   & 3    & 124  & 8   \\
HD 20807  & 1.0  & 198 & 36  & 1.1E-05 & 3.5E-06 & 2    & 1    & 10   & 3   \\
HD 21242  & 7.9  & 527 & 126 & 6.2E-04 & 4.5E-04 & 0.8  & 0.4  & 2    & 1   \\
HD 33262  & 1.5  & 132 & 35  & 1.8E-05 & 3.3E-06 & 6    & 4    & 27   & 20  \\
HD 35850  & 1.8  & 78  & 3   & 4.2E-05 & 2.6E-06 & 17   & 1    & 75   & 6   \\
HD 95698  & 8.6  & 211 & 31  & 2.5E-05 & 3.8E-06 & 5    & 1    & 13   & 3   \\
HD 95698  & 8.6  & 47  & 3   & 8.7E-05 & 3.9E-06 & 103  & 10   & 262  & 25  \\
HD 102647 & 13.6 & 118 & 1   & 2.2E-05 & 4.6E-07 & 21   & 1    & 45   & 1   \\
HD 113337 & 4.1  & 55  & 1   & 9.1E-05 & 2.7E-06 & 53   & 1    & 174  & 4   \\
HD 118216 & 21.0 & 211 & 16  & 1.8E-05 & 1.7E-06 & 8    & 1    & 15   & 2   \\
HD 119124 & 1.6  & 56  & 2   & 4.5E-05 & 4.2E-06 & 32   & 2    & 146  & 11  \\
HD 127762 & 33.4 & 85  & 4   & 9.7E-06 & 5.8E-07 & 62   & 6    & 99   & 10  \\
HD 131511 & 0.5  & 47  & 12  & 3.8E-06 & 8.9E-07 & 26   & 14   & 175  & 95  \\
HD 139006 & 71.4 & 124 & 2   & 1.3E-05 & 4.6E-07 & 42   & 2    & 52   & 2   \\
HD 150682 & 6.5  & 34  & 4   & 1.6E-05 & 2.3E-06 & 167  & 47   & 470  & 132 \\
HD 165908 & 2.9  & 45  & 4   & 1.6E-05 & 9.2E-07 & 64   & 11   & 241  & 40  \\
HD 172555 & 8.0  & 200 & 1   & 3.8E-04 & 2.7E-06 & 5.5  & 0.1  & 14.4 & 0.2 \\
HD 181296 & 21.6 & 134 & 2   & 2.1E-04 & 7.6E-06 & 20   & 1    & 37   & 1   \\
HD 196544 & 22.7 & 107 & 2   & 2.8E-05 & 7.2E-07 & 33   & 1    & 59   & 2   \\
HD 196885 & 2.7  & 24  & 7   & 2.8E-06 & 7.1E-07 & 226  & 135  & 868  & 517 \\
HD 202444 & 10.6 & 294 & 105 & 1.8E-05 & 1.8E-05 & 3    & 2    & 7    & 5   \\
HD 206860 & 1.2  & 90  & 7   & 8.5E-06 & 6.3E-07 & 11   & 1    & 55   & 7   \\
HD 207129 & 1.2  & 45  & 1   & 9.1E-05 & 3.2E-06 & 43   & 1    & 217  & 6   \\
HD 213398 & 37.2 & 122 & 21  & 1.4E-05 & 1.3E-06 & 31   & 9    & 47   & 14  \\
HD 213845 & 3.5  & 75  & 17  & 3.5E-06 & 4.5E-07 & 24   & 12   & 84   & 41  \\
HD 216956 & 16.6 & 69  & 3   & 7.6E-05 & 4.3E-06 & 66   & 7    & 134  & 14  \\
HD 223352 & 27.5 & 173 & 3   & 2.3E-05 & 7.3E-07 & 13.6 & 0.5  & 23   & 1   \\
HIP 8920  & 3.2  & 279 & 3   & 3.0E-02 & 7.8E-04 & 1.78 & 0.04 & 6.4  & 0.1 \\
HIP 11437 & 0.2  & 71  & 1   & 1.2E-03 & 4.0E-05 & 7.5  & 0.2  & 68   & 2   \\
HIP 17962 & 0.4  & 69  & 5   & 1.4E-04 & 2.6E-05 & 10   & 1    & 75   & 11 \\
\hline
\end{tabular}
\caption{Results of SED fitting for the 38 systems we find to have a significant infrared excess. The table shows the name of each system, its stellar luminosity, and the disc parameters $T_{\mathrm{disc}}$, $f$, $r_{\mathrm{bb}}$ and $r_{\mathrm{disc}}$ as defined in the text of section~\ref{sec:results} with their associated uncertainties. HD~95698 is listed twice because we model it with two dust components of different temperatures. A machine-readable version of this table is available online.}\label{tab:discparams}
\end{table*}

We find that 38 systems in our sample host debris discs, which are characterised by their fractional luminosity $f$ (i.e. the ratio of their luminosity to the stellar luminosity) and their temperature $T_{\mathrm{disc}}$. These parameters are listed in Table~\ref{tab:discparams} and plotted in Fig.~\ref{fig:fvsT}. The uncertainties shown in this Figure are calculated as half the difference between the 84th and 16th percentiles of the posterior distributions. The grey shading shows the region of parameter space in which we would not expect a disc around a typical star in our sample to be detectable by PACS at 70~$\mu$m. To estimate the detection threshold, we used equation~(8) of \citet{Wyatt08_Review}, taking the sensitivity to be 5~mJy\footnote{See Table~3.2 of the PACS Observer's Manual, \url{http://herschel.esac.esa.int/Docs/PACS/html/pacs_om.html}}, and the stellar distance and luminosity to be the median values in our sample: 23~pc and 2.1~$L_\odot$ respectively. Thus, it is not surprising that several systems lie slightly below this threshold, since the detection threshold for some stars will be lower than that of a median star. It is nonetheless a useful guideline, explaining the general form of the region where our discs lie, and why discs in certain parts of parameter space (i.e. near the threshold, where the detections have a lower signal to noise ratio) have comparatively large uncertainties. The points in Fig.~\ref{fig:fvsT} are coloured by stellar separation $a$, and from this it does not appear that $f$ or $T_{\mathrm{disc}}$ are strongly influenced by $a$ in the region where discs are detected.

\subsection{Dynamical stability}
\label{sec:stability}

Next, we wish to assess the dynamical stability of the discs. To do so, we need to establish where in the system they lie, i.e. their radii. We first convert the temperatures obtained from SED modelling into black body disc radii $r_{\mathrm{bb}}$ by assuming that the dust acts as a perfect absorber and emitter in thermal equilibrium with the stellar radiation (e.g. \citealt{Wyatt08_Review}). Radii calculated in this way, however, will underestimate the true radii of the discs by an amount that depends on the stellar luminosity $L$. The smallest grains in a disc are inefficient emitters, and so the more small grains that are present, the greater the departure from black body behaviour. The factor by which the black body radius underestimates the true radius is thus greater for lower luminosity stars, because brighter stars can remove larger grains via their radiation pressure (e.g. \citealt{Burns79_RadForces}). For Sun-like stars, this correction factor can be as large as $\sim$5 \mbox{(\citealt{Pawellek15_TrueRadius})}, which could make the difference between dust appearing to lie in a stable and an unstable region, and is thus important to account for. To estimate the true disc radii $r_{\mathrm{disc}}$, we use the prescription of \citet{Pawellek15_TrueRadius} -- specifically, we multiply our black body radii $r_{\mathrm{bb}}$ by the factor $\Gamma$ defined in their equation~(8), assuming that the dust is composed of a mixture of astrosilicate and ice. Note that in addition to the uncertainty arising from the need to assume a particular dust composition, there is some inherent scatter in the relation between $\Gamma$ and $L$, and so while we do on average expect the correction to improve our estimates of the disc radii there may be individual systems for which the black body radius is closer to the true value.

\begin{figure}
	\centering
    \hspace{-0.5cm}
	\includegraphics[width=0.5\textwidth]{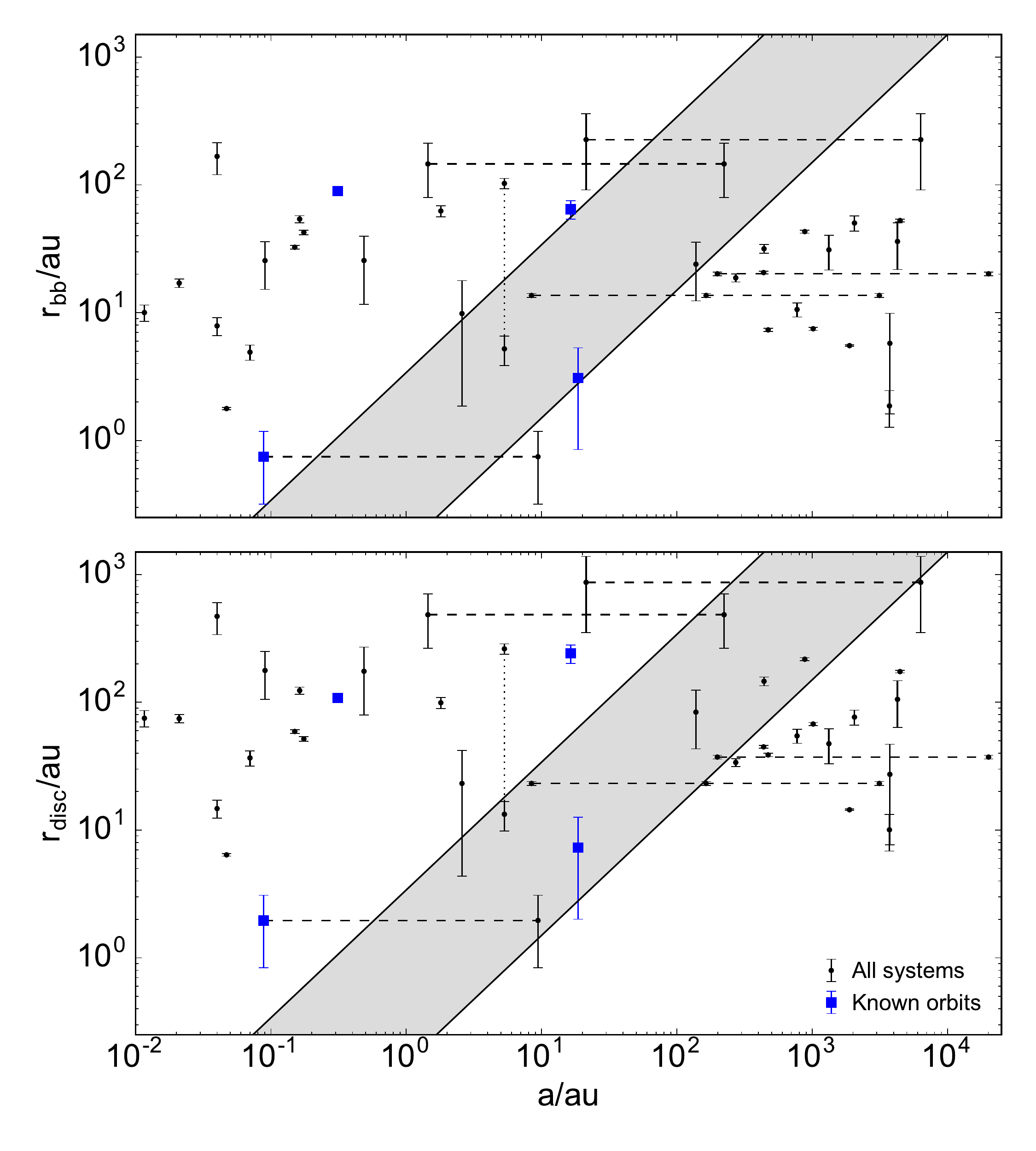}
	\caption{Disc radius versus stellar separation for our full sample. The upper panel shows black body disc radii, while the radii in the lower panel have been corrected for inefficient grain emission according to \citet{Pawellek15_TrueRadius}. Systems with more than two components, and hence more than one separation, are connected by dashed lines. We model HD~95698 as a two-temperature disc, and the two components are joined with a dotted line. HD~216956 (Fomalhaut) is the most widely separated system in our sample, and lies beyond the range of separations shown. Systems with highly graded VB6 orbits are plotted as large square points. The grey shading marks the approximate region of dynamical instability for planetesimals, assuming an equal-mass binary with an eccentricity of 0.4, as in \citet{Rodriguez15_BinaryDebris}. Around 24\% of the discs we find appear to be unstable based on the lower panel.}
	\label{fig:rvsa}
\end{figure}

Both $r_{\mathrm{bb}}$ and $r_{\mathrm{disc}}$ are listed in Table~\ref{tab:discparams} and plotted as functions of stellar separation in Fig.~\ref{fig:rvsa}. Also shown in this Figure, with grey shading, is the approximate dynamically unstable region of parameter space as calculated by \citet{Rodriguez15_BinaryDebris}, who used the results of \citet{Holman99_Stability} and assumed equal stellar masses and an orbital eccentricity of 0.4. The boundaries of this region are at 0.15 and 3.4 times the stellar separation. Note that a system lying in this region can only be considered definitively unstable if the value of $a$ used is the true semimajor axis of the stellar orbit. As discussed in section~\ref{sec:sample}, this is not in fact the case for most systems in our sample. For visual binaries -- i.e. those towards the right hand side of Fig.~\ref{fig:rvsa} -- generally only the \textit{projected} separation is known, and this is used for $a$. In contrast, the $a$ values for spectroscopic binaries, which lie towards the left hand side, will have been derived from known periods. While the conversion from period to semimajor axis does require an estimate of the total stellar mass $M$, this can be established from stellar spectra sufficiently accurately for our purposes (up to a factor $\sim$2 for equal mass binaries), and the dependence on $M$ is not strong ($a\propto M^{1/3}$), which means that semimajor axis estimates for spectroscopic binaries will generally be reliable. Thus, while in all but a few cases $a$ is not a direct measurement of the semimajor axis, systems lying further towards the left hand side of the plot will generally have an $a$ value closer to their true semimajor axis.

Of the 38 discs we find, 9 have an $r_{\mathrm{disc}}$ placing them in the nominally unstable region: HD~19994, HD~21242, HD~95698, HD~119124, HD~181296, HD~202444, HD~207129, HD~213845 and HD~223352. We generally expect $r_{\mathrm{disc}}$ to give a better estimate of the disc radii, and thus a better indication of stability, than $r_{\mathrm{bb}}$. However, to allow a direct comparison with previous work we note that if the black body radii \textit{were} to be interpreted as true disc radii, then we would conclude that the only unstable systems were HD~19994, HD~95698, HD~202444, HD~213845 and HD~223352. That is, correcting for inefficient grain emission moves four systems up into the unstable region, without moving the other five far enough out to be considered stable. 

Assuming binomial uncertainties (e.g. \citealt{Burgasser03_BinomialErrors}), the nine potentially unstable systems constitute $24^{+8}_{-6}\%$ of the discs. For comparison, of the 71 systems considered to host discs in \citet{Trilling07_BinaryDebris}, \citet{Rodriguez12_BinaryDebris} and \citet{Rodriguez15_BinaryDebris} combined, a total of 13 were concluded to be unstable, or $18^{+5}_{-3}\%$. While our derived apparent instability fraction \textit{is} therefore consistent with the combined results of these three studies, our specific set of unstable discs is quite different. In fact, in many cases we do not detect a disc in the previously proposed unstable systems at all, which emphasises the importance of revisiting the conclusions of previous work. For example, none of the three systems with `unstable discs' from \citet{Trilling07_BinaryDebris} -- HD~46273, HD~80671 and HD~127726 -- in fact satisfy our excess criterion. Note that while a MIPS 70~$\mu$m observation of HD~127726 exists from \citet{Trilling07_BinaryDebris}, we do not include the corresponding flux in our SED fitting because it is 4.4$\sigma$ (or a factor of $\sim$3) above the \texttt{OT2_gkennedy_2} PACS 70~$\mu$m flux, probably due to contamination from a background source which is clearly visible in the PACS image $\sim$20$^{\prime\prime}$ away; this is likely the reason for our different conclusions for this particular system. Similarly, none of the three `unstable' systems from \citet{Rodriguez12_BinaryDebris} that are in our sample (HD~56986, HD~73752 and HD~174429) appear to have a significant excess. Of the four `unstable' \citet{Rodriguez15_BinaryDebris} systems, our results agree that HD~223352 and HD~19994 do seem to host unstable dust, but we conclude that the disc of HD~165908 (99 Her) is in a stable location based on either $r_{\mathrm{disc}}$ or $r_{\mathrm{bb}}$ (see also \citealt{Kennedy12_99Her}) and that~HD 133640 does not have a significant infrared excess. For reference, the far infrared significances of the systems previously thought to host an unstable disc which we do not consider to have an excess are displayed in Table~\ref{tab:prev_unstable_excess}. 

\begin{table}
\begin{tabular}{lccc}
\hline

 & \multicolumn{3}{c}{$\chi$} \\
 \cline{2-4} \\
Name & MIPS 70~$\mu$m & PACS 70~$\mu$m & PACS 100~$\mu$m \\
\hline
HD 46273  & 2.52        & -0.35       &              \\
HD 56986  & -0.62       &             & 1.06         \\
HD 73752  & 0.15        &             & -0.06        \\
HD 80671  & 1.91        & 1.08        &              \\
HD 127726 &             & -0.52       &              \\
HD 133640 & 3.74        &             & 2.03         \\
HD 174429 & 0.62        & 1.78        & 1.02         \\
\hline
\end{tabular}
\caption{Far infrared significances of the seven systems previously considered to host unstable dust and for which we do not detect an excess. Recall that the excess thresholds are set by Figs~\ref{fig:chihist_nodrake} and \ref{fig:chihist_drake}, and can thus exceed $\chi=3$. In particular, the MIPS 70~$\mu$m threshold is $\chi=5.00$.}
\label{tab:prev_unstable_excess}
\end{table}

Fig.~\ref{fig:rvsa} also highlights (as larger, square points) those systems for which the stellar orbits can be considered well known, i.e. those listed in VB6 with grade 1 or 2, as discussed in section~\ref{sec:sample}. These are the systems for which the plotted stellar separations \textit{are} true semimajor axes, which allows a stronger conclusion to be drawn about their discs' stability than if only the projected separation were known. Recall also that spectroscopic binaries are not subject to projection effects, which in practice means that although they do not have directly measured semimajor axes like the highly graded VB6 systems, the horizontal positions of most binaries with $a$ less than around 1~au in Fig.~\ref{fig:rvsa} are well constrained, allowing us to conclude that such systems are very likely stable.

The disc-bearing known VB6 orbit systems are HD~13161 (beta Tri), HD~165908 (99 Her), HD~202444 (tau Cygni), and HD~21242 (UX Ari). The first two of these systems are well known from previous work to host discs (\citealt{Kennedy12_99Her}, \citeyear{Kennedy12_CoplanarCircumbinary}), and lie well outside the unstable zone. The third, HD~202444, is a new detection from the \texttt{OT2_gkennedy_2} survey, and lies in the unstable zone in Fig.~\ref{fig:rvsa}. Finally, HD~21242, one of the \texttt{OT1_jdrake01_1} targets, is a triple system, and it is only the AaAb component which has a known orbit. Fig.~\ref{fig:rvsa} shows that the dust is sufficiently far out to be stable to perturbations from the AaAb pair, though it does lie just inside the nominally unstable region of the less well constrained AB pair. 

\citet{Rodriguez15_BinaryDebris} noted that two possible effects which could cause stable discs to \textit{appear} unstable in a plot of $r_{\mathrm{disc}}$ against $a$ are inefficient grain emission (i.e. the black body disc radius underestimating its true radius) and orbital projection (i.e. lack of knowledge of the true stellar semimajor axis). We have corrected for the former effect to the extent that this is possible and still found that a significant fraction of discs lie in the unstable region. In addition, while the number of disc-bearing binaries whose semimajor axes are well known is very limited, we found that one out of the four such discs in our sample appears unstable.

\subsection{Assessment of potentially unstable systems}
\label{sec:unstablesys}

In the previous subsection, we identified nine systems which, based on their values of $r_{\mathrm{disc}}$, appear to host unstable dust. As these systems are of particular interest, here we discuss each of them individually and assess the evidence that their discs truly are unstable.

\begin{figure}
	\centering
	\hspace{-0.5cm}
	\includegraphics[width=0.5\textwidth]{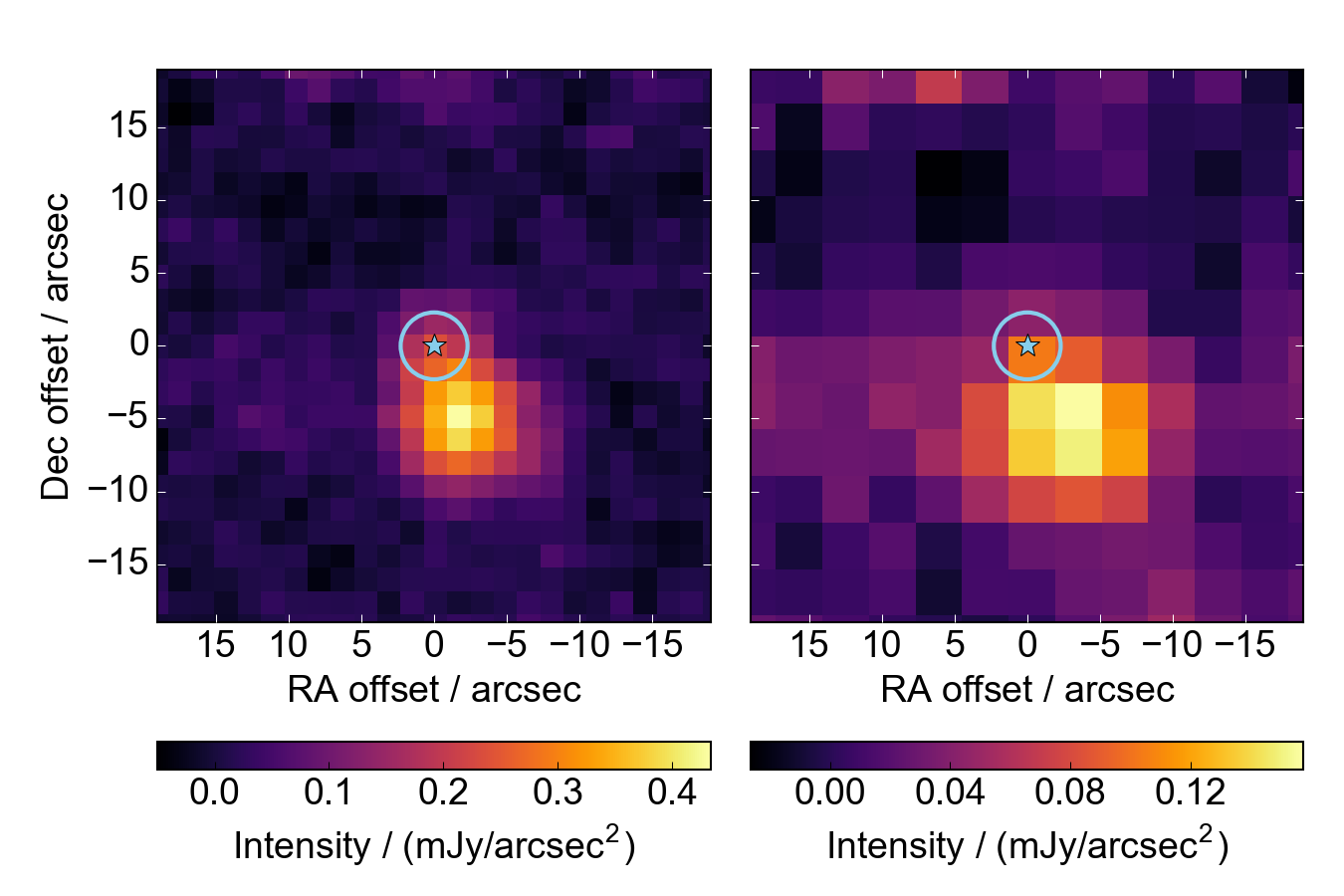}
	\caption{Cutouts of the PACS images of HD~19994, at 100~$\mu$m (left) and 160~$\mu$m (right). The expected stellar position at the time of observation is marked, along with an error circle reflecting the \textit{Herschel} pointing uncertainty of 2.3$^{\prime\prime}$. The 100~$\mu$m image appears to show two sources: one at the expected stellar position, and another, brighter source around 5$^{\prime\prime}$ away. In the 160~$\mu$m image, only the brighter of the two sources is visible. This may suggest that the image is contaminated by a background galaxy. }
	\label{fig:hd19994}
\end{figure}

\textit{HD~19994} -- this is a triple star system; the B and C components are on a close orbit ($a\sim1.5~\mathrm{au}$, from VB6), while the BC pair is at a projected separation of $a\sim220~\mathrm{au}$ from the A component (\citealt{Rodriguez15_BinaryDebris}). We derive a disc radius of $480\pm220~\mathrm{au}$. The simplified picture presented by Fig.~\ref{fig:rvsa} suggests that the disc is stable to perturbations from the BC pair but formally lies just inside the unstable region of the A-BC system. \citet{Wiegert16_94Ceti} studied HD~19994 in detail, and using radiative transfer modelling and $N$-body simulations of different possible disc geometries, they concluded that the disc, if real, is likely circumsecondary, i.e. orbiting the BC pair, and in a stable location. They also noted that there is a 2.3\% probability that the 100~$\mu$m excess is due to contamination from a background galaxy. This contamination probability depends on wavelength and PSF size, and so its exact value will vary within our sample depending on the data that exist for each star. However, given our sample size of several hundred systems, assuming a value on the order of 1\% it is expected that several stars in our sample will be affected by background galaxies. For this system in particular, the PACS images, shown in Fig.~\ref{fig:hd19994}, do hint at background contamination: at 100~$\mu$m there appears to be one source very close to the expected stellar position at the time of observation, with another, brighter source around 5$^{\prime\prime}$ away. At 160~$\mu$m, only the latter of these appears to be detected. Note that in Fig.~\ref{fig:hd19994}, the stellar position shown is calculated using \textit{Hipparcos} proper motions; the error circle around the star has radius 2.3$^{\prime\prime}$, reflecting the \textit{Herschel} pointing uncertainty\footnote{See Table~2.1 of the \textit{Herschel} Observer's Manual, \url{http://herschel.esac.esa.int/Docs/Herschel/html/Observatory.html}}. PSF fitting favours subtracting the brighter source, but it is possible that this is in fact a galaxy while the less bright source is the star. In summary, we do not consider the evidence that this system genuinely contains unstable dust, or perhaps a disc at all, to be compelling.

\textit{HD~21242 (UX Ari)} -- another triple system, with AaAb on a close orbit and a more distant B component. Estimating the AaAb and AB semimajor axes from SB9 periods as outlined in section~\ref{sec:sample} gives 0.08 and 9~au respectively. The periods we used are those calculated by \citet{Duemmler01_UXAri} using radial velocity observations; the AB period is based on a fit to the velocity of the centre of mass of the AaAb pair, which they found to be changing systematically with time. The AaAb orbit is listed with grade 2 in VB6, giving a semimajor axis of 0.09~au, very close to the value we inferred from the period. AB also appears in VB6 with a semimajor axis of 33~au; as this is graded 5, we take the AB semimajor axis to be 9~au. Note that the AB component lies very close to the edge of the nominally unstable region in Fig.~\ref{fig:rvsa}, and using the VB6 value would move the system well into the stable region. We find a disc at $2\pm1~\mathrm{au}$. It is the hottest disc in our sample by some margin ($T_{\mathrm{disc}}\sim500~\mathrm{K}$), and the third brightest ($f\sim6\times10^{-4}$), making it somewhat exceptional. \citet{Matranga10_CloseBinaries} similarly identified this system as hosting large amounts of very warm dust, and suggested planetary collisions as a potential origin of this dust. We discuss this idea further in section~\ref{sec:closebinaries}. If this system hosts dust which is truly residing in the unstable zone, then binary perturbations from the AB pair could provide an explanation for the destabilisation of the planets proposed to be responsible. Alternatively, note that the dust is consistent with lying at the boundary of instability (which is formally at 1.4~au); this could imply that the disc is being truncated at its outer edge by perturbations from the B component.

\textit{HD~95698} -- a binary with projected separation 5~au (\citealt{Trilling07_BinaryDebris}). As noted in section~\ref{sec:SED_modelling}, inspection of its IRS spectrum motivated us to model this system with two temperature components. This suggests that the disc likely has significant radial extension and/or multiple belts, though two-temperature discs can also arise from a single narrow belt, in which different grain sizes have different temperatures (\citealt{Kennedy14_TwoTemp}). While the cooler component lies well into the stable zone (at $262\pm25~\mathrm{au}$), the warmer component is formally unstable with $r_{\mathrm{disc}}=13\pm3~\mathrm{au}$. Given that the boundary of the grey region in Fig.~\ref{fig:rvsa} is blurred by the assumption of a fixed stellar mass ratio and eccentricity, and that the warmer component here lies very close to this boundary (which is at 18~au), the implication may be that the dust responsible for the warm emission is being truncated at its inner edge by binary perturbations. This disc is also resolved at 70~$\mu$m in the \texttt{OT2_gkennedy_2} observations, and the details of our image modelling are presented in appendix~\ref{app:imagemodel}. Our conclusions from this modelling are similar to our findings from SED fitting: the image is consistent with a broad disc extending from $101^{+72}_{-38}~\mathrm{au}$ to $318^{+43}_{-54}~\mathrm{au}$, corresponding to the cooler component of our SED model. We also allow for a component of unresolved flux, and find that the best-fitting model favours such a component being present at 3~mJy (in addition to the $\sim$5~mJy from the star), around 2.6\% as bright as the resolved component; the warmer component of our SED model is likely responsible for this emission. In our SED fit the warmer component similarly has 2.7\% of the flux of the cooler component at 70~$\mu$m. As the warmer component is not resolved, the image cannot give a better constraint on its radius than SED modelling, and our conclusion about its stability is not affected.

\textit{HD~119124} -- another binary, with projected separation $a\sim440~\mathrm{au}$ (\citealt{Rodriguez12_BinaryDebris}). The excess is unambiguous, with a 14$\sigma$ detection in MIPS 70~$\mu$m. It has not been previously identified as unstable, and if we had taken the disc radius to be equal to the black body radius of $32\pm2~\mathrm{au}$ then we would also have concluded that it is stable. As a late F-type star, its stellar luminosity correction factor is large (\citealt{Pawellek15_TrueRadius}), placing its estimated `true' disc radius firmly in the unstable region, with $r_{\mathrm{disc}}=145\pm11~\mathrm{au}$. However, note that the quoted uncertainty of 11~au does not acknowledge the uncertainty inherent in the correction factor, so the disc could be stable if the dust composition is different to that assumed.

\textit{HD~181296 (eta Tel)} -- a triple system, with the secondary companion $200~\mathrm{au}$ away (\citealt{Rodriguez12_BinaryDebris}) in projected separation, and a much more distant tertiary companion at $\sim2\times10^4~\mathrm{au}$ (\citealt{Tokovinin18_MSC}), sufficiently distant that it will not affect the stability of the circumprimary disc. The tertiary companion is HD~181327, which hosts its own debris disc (\citealt{Lebreton12_HD181327}), but as we are focusing on observations of primary stars, we plot both components at our calculated circumprimary disc radius in Fig.~\ref{fig:rvsa}. We find $r_{\mathrm{bb}}=20\pm1~\mathrm{au}$ and $r_{\mathrm{disc}}=37\pm1~\mathrm{au}$. Mid infrared images of HD~181296 have revealed a solar system-like architecture, with a resolved disc at $24\pm8~\mathrm{au}$ and an unresolved component at $\sim4~\mathrm{au}$ (\citealt{Smith09_etaTel}); thus, in this case, the black body radius actually provides a slightly better estimate of the true radius. Based on the resolved radius, the disc is \textit{not} unstable. Note also that the primary is an early A type star, while the secondary companion is in fact a brown dwarf (\citealt{Neuhauser11_etaTel}), meaning that the mass ratio here is particularly extreme. The grey region in Fig.~\ref{fig:rvsa} assumes equal masses for both stars, so the unstable region for this particular system will be narrower than the Figure suggests.

\textit{HD~202444} -- this binary system has a semimajor axis of $a=18.6~\mathrm{au}$, from VB6. This is the only potentially unstable system whose orbit is well known. The PACS 70~$\mu$m data from \texttt{OT2_gkennedy_2} show an excess with significance $\chi=4.54$. As this is a single-band excess whose signal to noise ratio is not particularly high, the disc radius is not well constrained, with $r_{\mathrm{disc}}=7\pm5~\mathrm{au}$. Given that the excess is detected by PACS at 70~$\mu$m but not 160~$\mu$m, by considering the detection limits as a function of disc radius in these two bands we can deduce an upper limit on $r_{\mathrm{bb}}$ of $\sim$100~au, corresponding to an $r_{\mathrm{disc}}$ of $\sim$240~au. Note that this upper limit appears incompatible with the above quoted disc radius and its uncertainty. This is likely a result of the fact that when a model with only a stellar component is fitted to the system's SED, the \textit{AKARI} (\citealt{Ishihara10_AKARI}) 9~$\mu$m flux is $\sim$4$\sigma$ below the photosphere, while the MIPS 24~$\mu$m flux is $\sim$2$\sigma$ above it. Adding a dust component to the model allows the residuals in both of these bands to be made less significant: the stellar component preferentially has a lower flux than in the star-only model, set by the \textit{AKARI}  9~$\mu$m photometry, while the MIPS 24~$\mu$m flux -- which is now even further above the photosphere, at $\sim$7$\sigma$ -- forces the dust to a relatively high temperature of $\sim$300~K. It could be the case that the \textit{AKARI} photometry is subject to underestimated uncertainties and/or a systematic effect, either of which could make the low flux measured for this system unreliable. If so, the MIPS 24~$\mu$m excess would not be so significant and the dust could be much cooler, as indicated by the upper limit derived only from consideration of the PACS sensitivities. Despite the poorly constrained radius, as orbital projection is not an issue for this system it may be an interesting target for future observations.

\textit{HD~207129} -- a binary system with projected separation $a\sim880~\mathrm{au}$ (\citealt{Rodriguez15_BinaryDebris}). Its ring-like disc has been resolved both in scattered light, with a measured radius of around 160~au (\citealt{Krist10_HD207129}), and by \textit{Herschel}, with a measured radius of $140\pm30~\mathrm{au}$ (\citealt{Marshall11_HD207129}). Our SED modelling gives a black body radius of $43\pm1~\mathrm{au}$; as a G-type star, its luminosity-corrected radius is considerably larger, at ${r_{\mathrm{disc}}=217\pm6~\mathrm{au}}$. Thus, in this case the corrected radius does indeed allow a better assessment of stability, since it overestimates the \textit{Herschel}-resolved radius by a factor of only 1.6, while the black body radius is around 3.3 times too small. The lower boundary of the formally unstable region at a semimajor axis of 880~au is at $\sim$130~au, and so based solely on $r_{\mathrm{disc}}$ the dust would be considered unstable. However, the \textit{Herschel}-resolved radius, which directly measures the true disc size, while still formally unstable, is extremely close to -- and in fact consistent with -- the boundary. It may thus be the case that the disc is not truly unstable, and either the binary semimajor axis is greater than the projected separation or the disc is being truncated at its outer edge by the binary companion.

\textit{HD~213845} -- a binary with projected separation $a\sim140~\mathrm{au}$ (\citealt{Rodriguez15_BinaryDebris}) and ${r_{\mathrm{disc}}=80\pm40~\mathrm{au}}$. It has a marginal excess in PACS 100~$\mu$m, with $\chi=3.88$, hence the relatively poorly constrained radius. Its MIPS 70~$\mu$m flux is notably high too, but still below our threshold, with $\chi=3.20$. As with HD~119124, the system lies well into the apparently unstable region in Fig.~\ref{fig:rvsa}, and there is no obvious reason to disbelieve the excess. The PACS 100~$\mu$m image is not resolved; while the image has a low signal to noise ratio, this may suggest that the true disc radius is smaller than 80~au. A disc of this size would have an angular diameter of 7.0$^{\prime\prime}$ at this system's distance of 22.7~pc, somewhat larger than the PACS beam size of $6.65^{\prime\prime}\times6.87^{\prime\prime}$ (\citealt{Poglitsch10_PACS}).

\textit{HD~223352} -- this is a hierarchical quadruple system. The AB separation is $160~\mathrm{au}$, with the B component itself being a binary of separation $\sim8~\mathrm{au}$; the C component (HD~223340) is seen at a projected separation of $\sim3100~\mathrm{au}$ from the AB system (\citealt{Rodriguez15_BinaryDebris}). As noted in that paper, the system is unusual in that \textit{two} of its constituent stars -- A and C -- are known to host their own debris discs (\citealt{Phillips2011_DDSurveys}). Fig.~\ref{fig:rvsa} is made without any knowledge of which star the detected dust orbits, and so we plot all three separations listed above at the same dust radius: ${r_{\mathrm{bb}}=13.6\pm0.5~\mathrm{au}}$ or ${r_{\mathrm{disc}}=23\pm1~\mathrm{au}}$. In fact, inspection of Fig.~\ref{fig:rvsa} shows that in our study this system only appears unstable if one plots the close BaBb separation -- however, given the knowledge that the dust is not orbiting the B pair, it becomes clear that this separation is not relevant. The dust \textit{is} stable to perturbations from the C component, and is formally stable -- but at the boundary ($\sim24~\mathrm{au}$) -- with respect to the B component. The black body radius for the circumprimary disc derived by \citet{Rodriguez15_BinaryDebris} is somewhat larger than our value, at 27~au, which led them to the similar conclusion that the dust is at the boundary of stability (though formally unstable). This could be another case in which either the circumprimary disc is being truncated at its outer edge by perturbations from the B component or the true binary semimajor axis is greater than the projected stellar separation.

We now summarise our findings from the above consideration of individual systems. The main conclusion here is that while some systems do appear based on Fig.~\ref{fig:rvsa} to host unstable discs, a more detailed consideration of these systems highlights the fact that this result should not be taken at face value. For one system (HD~19994) we are doubtful about the existence of a disc, due to possible background contamination. For another system (HD~223352), the disc only appears unstable if we consider the BaBb separation, but from previous work it is in fact known to be circumprimary. There is also a system (HD~181296) whose disc radius from resolved image modelling differs from our radius from SED fitting and indicates that the disc is actually stable. 

For two systems (HD~21242 and HD~95698) our calculated radii place them very close to the boundary of stability, as does the resolved radius of another (HD~207129). In such cases, we are particularly hesitant to conclude that the discs are genuinely unstable. This is because the boundary of instability is not as well defined as it appears in Fig.~\ref{fig:rvsa}, since the mass ratio and eccentricity vary between systems and are not generally well known. Given the proximity of these systems to the nominal boundary, it could be the case that these discs are stable and being truncated by binary perturbations; however, uncertainty on the true positions of systems in Fig.~\ref{fig:rvsa} (as discussed in detail below) means that the evidence for the truncation hypothesis is not strong. We note, though, that it is known that discs undergoing truncation could appear unstable: \citet{Thebault10_BinaryDiscs} found using $N$-body simulations that a circumprimary disc being truncated at its outer edge by a binary companion will extend into the dynamically unstable region to some degree, as a result of collisionally-produced small grains which are placed on eccentric orbits by radiation pressure. This can alter the SED of a system with such a disc sufficiently to push its inferred radius into the unstable region, even though the parent planetesimals are stable. 

The three remaining systems, which lie well into the unstable region in Fig.~\ref{fig:rvsa}, are HD~119124, HD~213845 and HD~202444. These may be interesting targets for future observations, however there are still effects which make their true positions in that Figure uncertain. We found by considering the resolved radius of HD~181296 that the `corrected' disc radius $r_{\mathrm{disc}}$ is not always a better indicator of true disc size than the black body radius; this is not surprising given that the correction requires an assumption of dust composition and that the correction factor is known to show some inherent scatter (\citealt{Pawellek15_TrueRadius}). Thus, the vertical positions of the systems in Fig.~\ref{fig:rvsa} are actually less well known than the error bars suggest. We note that the black body radii $r_{\mathrm{bb}}$ of HD~213845 and HD~202444 lie at the boundary of instability, and that of HD~119124 lies in the stable region.

\begin{figure}
	\centering
    \hspace{-0.5cm}
	\includegraphics[width=0.5\textwidth]{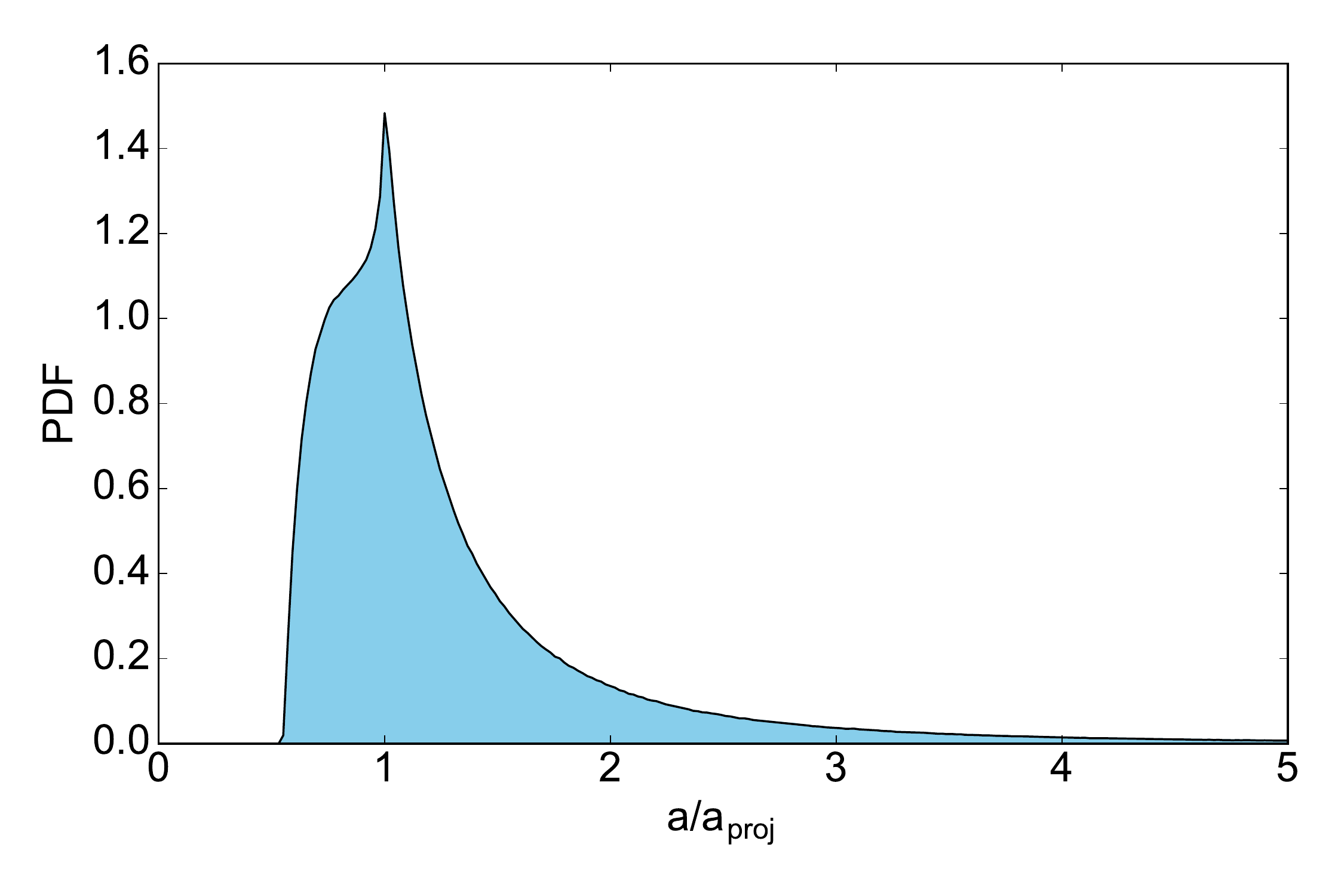}
	\caption{Probability density function (PDF) of the ratio of true semimajor axis to projected separation for a population of binaries with uniformly distributed orbital elements. Eccentrities were taken to be distributed between 0 and 0.8.}
	\label{fig:projection}
\end{figure}

There is also the possibility of orbital projection for visual binaries introducing uncertainty on the horizontal positions. To quantify the importance of this effect, we generated $10^7$ sets of random uniformly distributed orbital elements, with inclination $I$ ranging from 0 to $\pi/2$, eccentricity $e$ from 0 to 0.8, argument of pericentre $\omega$ from 0 to $2\pi$, and mean anomaly $M$ from 0 to $2\pi$. The eccentricity distribution we used is an approximation to what has been observed in a real population of binaries (see Fig.~14 of \citealt{Raghavan10_Multiplicity}). We used uniformly distributed mean anomalies rather than true anomalies so that the fact that the stars spend more time near apocentre than near pericentre is accounted for. For each set of elements, we calculated the ratio of the true semimajor axis $a$ to the projected separation $a_{\mathrm{proj}}$, using equation~(2.122) of \citet{Murray1998}. The resulting distribution of $a/a_{\mathrm{proj}}$ is shown in Fig.~\ref{fig:projection}. It is clear that the distribution is highly skewed, and while the most probable result \textit{is} that the semimajor axis is equal to the observed projected separation, there is a non-negligible probability that the true value is several times larger. The median value of $a/a_{\mathrm{proj}}$ is $1.06^{+0.69}_{-0.29}$, where the quoted uncertainties are based on the 16th and 84th percentiles of the distribution; its mean is 1.39. Given the projected separations of HD~119124 and HD~213845, and assuming no uncertainty on their luminosity-corrected disc radii, the distribution in Fig.~\ref{fig:projection} implies that the probabilities that the true semimajor axes of these systems lie outside the grey region in Fig.~\ref{fig:rvsa} are 9.8\% and 2.9\% respectively. The actual probabilities of stability will be even larger, since, as discussed above, the vertical positions of the points are uncertain.

Given the expected short lifetime of unstable dust, we consider it likely that some combination of projection and uncertainty on the true disc radius is responsible for the apparent instability of HD~119124, HD~213845 and HD~202444. Recall also that \citet{Trilling07_BinaryDebris} suggested the inward migration via PR drag of dust produced in a more distant stable planetesimal belt as an origin for unstable dust in binary systems. Aside from the fact that dust is expected to collisionally evolve faster than PR drag can act (as discussed in section~\ref{sec:intro}), there is the issue that HD~119124, HD~213845 and HD~202444 all have $r_{\mathrm{disc}}<a$ and thus appear to be circumstellar rather than circumbinary, meaning that there are no suitable stable locations for planetesimals from which the dust could drift inwards. We finish by highlighting HD~202444 as perhaps the best candidate for a genuinely unstable system, since it has a known orbit and is thus not subject to horizontal uncertainty, with the caveat that its infrared excess is detected at relatively low signal to noise.

\subsection{Disc statistics}
\label{sec:statistics}

\begin{figure}
	\centering
    \hspace{-0.5cm}
	\includegraphics[width=0.5\textwidth]{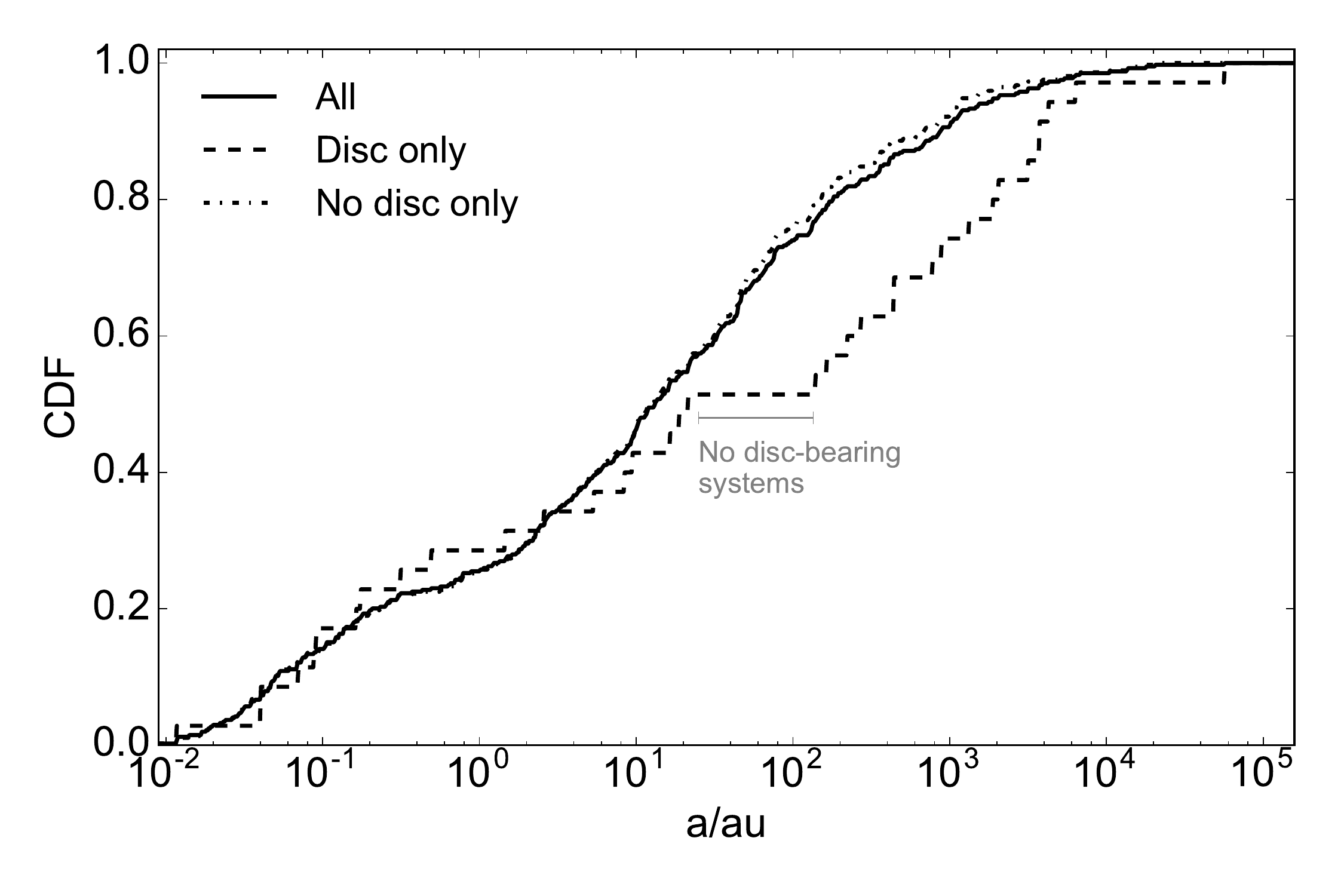}
	\caption{Cumulative distribution functions (CDFs) of stellar separations, for all unbiased systems in our sample, for disc-bearing systems, and for systems without a detectable disc. There is an absence of disc-bearing systems with separations in the range $\sim$25--135~au. The grey shading highlights the separation range for which discs are lacking.}
	\label{fig:sma_CDF_all}
\end{figure}

Having identified and characterised the discs in our sample, we now turn our attention to the issue of whether the likelihood of finding a detectable level of dust in a binary or multiple system depends on its orbit. For the purposes of this subsection, we exclude the 13 systems which we took from \citet{Rodriguez12_BinaryDebris} and which do \textit{not} also appear in the samples of \citet{Trilling07_BinaryDebris} or \citet{Rodriguez15_BinaryDebris}. This is because all systems compiled by \citet{Rodriguez12_BinaryDebris} were included in their study specifically because they were thought from previous work to host debris discs. In contrast, all other sources from which we compile our complete sample are unbiased towards the detection of discs. Thus, including the \citet{Rodriguez12_BinaryDebris} systems could affect our statistical conclusions, making discs appear to be more common than they really are. Note that although seven of the \texttt{OT2_gkennedy_2} systems were included in that programme to follow up on the \citet{Trilling07_BinaryDebris} results (see section~\ref{sec:sample}) -- and are therefore part of the \texttt{OT2_gkennedy_2} sample specifically because they have discs -- these systems fall in the unbiased \citet{Trilling07_BinaryDebris} sample anyway and are thus not excluded from the statistical analysis. The resulting `statistical' sample thus contains 328 systems, with a total of 403 stellar separations.

In Fig.~\ref{fig:sma_CDF_all}, we plot the cumulative distribution function (CDF) of stellar separations for the set of all unbiased systems, as well as separate CDFs for the disc-bearing subset and the subset for which no disc is detected. As in Fig.~\ref{fig:sma_dist_per_sptype}, systems of more than two stars are counted multiple times (once at each individual separation). The most striking feature of this plot is that the CDF for disc-bearing systems is constant between around 25 and 135~au -- i.e. no discs are detected in systems with separations in that range. The results of \citet{Rodriguez15_BinaryDebris} similarly hinted that discs are less common in binaries of intermediate separation than in close or wide systems. However, on performing a KS test on the separations of their disc-bearing and disc-free systems, they obtained a $p$-value of $0.09$, and were thus unable to conclude that the two distributions were in fact different (assuming that a confidence level of 95\% is required to draw such a conclusion). We performed the equivalent test on our CDFs, yielding a $p$-value of $0.006$. This indicates that the separations of the disc-bearing and disc-free systems are indeed drawn from different distributions, at a confidence level of 99.4\%. We therefore conclude that detectable cool debris discs are less common in binary systems with separations between $\sim$25 and 135~au.

\begin{figure}
	\centering
    \hspace{-0.5cm}
	\includegraphics[width=0.5\textwidth]{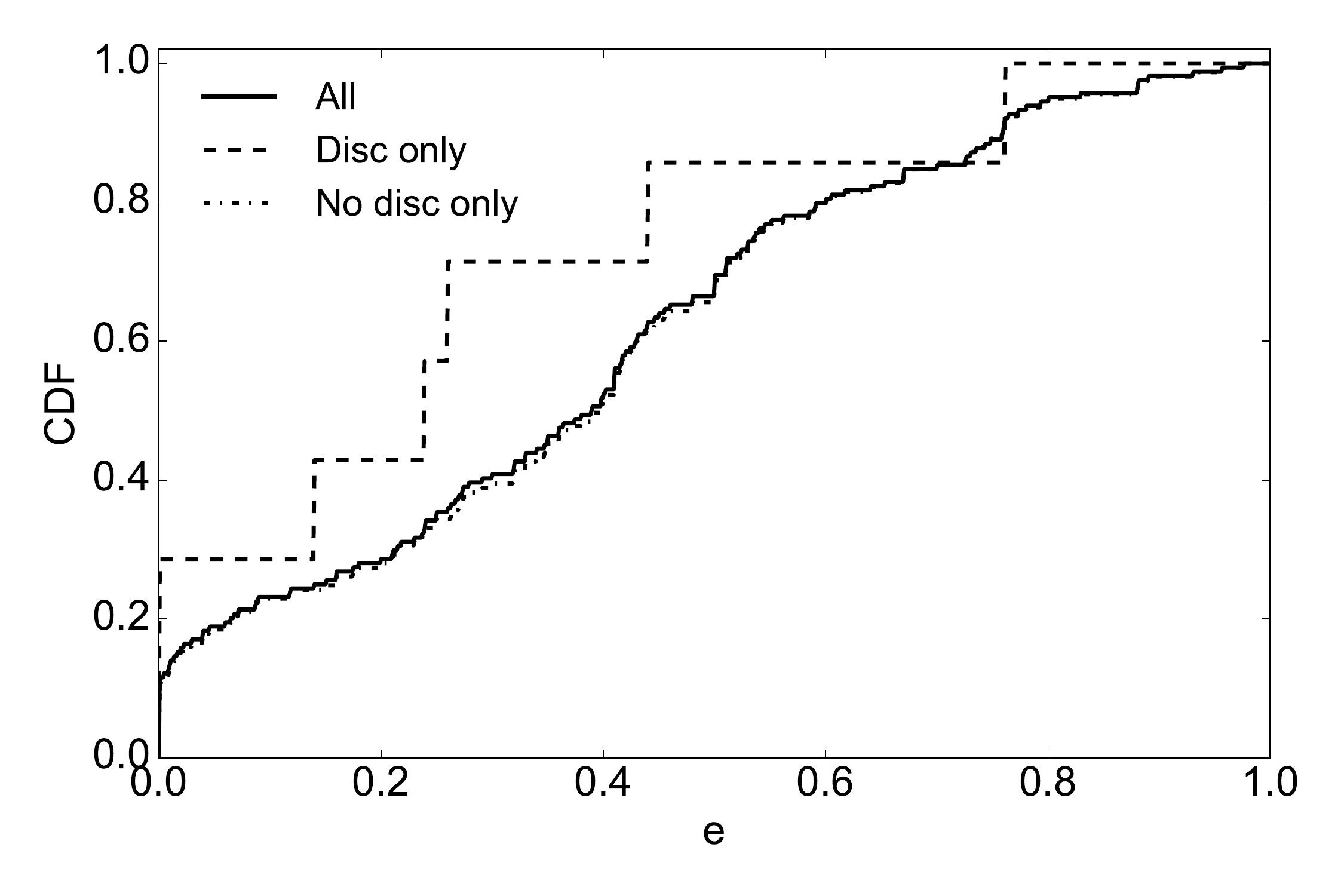}
	\caption{Cumulative distribution functions (CDFs) of orbital eccentricities, for all unbiased systems in our sample with eccentricity information, for disc-bearing systems, and for systems without a detectable disc. The CDFs are statistically indistinguishable.}
	\label{fig:ecc_CDF_all}
\end{figure}

We also performed a KS test on the eccentricities $e$ of the disc-bearing and disc-free systems and found no evidence for a significant difference between their distributions, with a $p$-value of $0.86$ if only systems with highly graded VB6 eccentricities are included, or $0.25$ for systems with \textit{any} available eccentricity (either in VB6 with any grade or from the literature). This may suggest that the eccentricity of a binary does not significantly affect its likelihood of hosting detectable dust, but it should also be noted that we only have eccentricity information for 139 out of 341 systems. These systems are strongly biased towards having close separations, with 92\% of the known eccentricities corresponding to separations of less than 100~au. The relatively large $p$-values do not necessarily rule out a dependence of dust detectability on binary eccentricity, but could simply result from the relatively small numbers involved. Eccentricity CDFs for all unbiased systems with eccentricity information, and for disc-bearing and disc-free systems separately, are shown in Fig.~\ref{fig:ecc_CDF_all}. These CDFs hint at the result that disc-bearing systems tend to have lower eccentricities than disc-free systems, or equivalently that the occurrence rate of detectable discs is higher for lower eccentricity binaries. If this is a genuine physical effect, it could be a result of the fact that more eccentric binaries will excite orbiting planetesimals to higher eccentricities, thus increasing their relative velocities and reducing their collisional lifetime.

While the general conclusion that debris discs in intermediate separation binaries are somewhat less common is not unexpected based on previous work, we were surprised to find a complete absence of discs over an order of magnitude of separations. Note that \citet{Phillips2011_DDSurveys} similarly found an absence of detectable debris discs in binaries with separations of $\sim$3--150~au using MIPS photometry, though for a smaller sample of $\sim$50 systems, all of which were A-type. As a check on our result, we inspected the SEDs of systems with separations between 25 and 135~au which previous work concluded \textit{did} have a disc (listed below) individually. From \citet{Rodriguez15_BinaryDebris}, HD~133640 has both MIPS 70~$\mu$m and PACS 100~$\mu$m photometry, neither of which we deem significant; HD~202940 has photometry in these bands as well as PACS 70~$\mu$m, and again none are significantly in excess, though the PACS 100 and 160~$\mu$m fluxes are both at $3.3\sigma$. Infrared excesses peaking beyond 100~$\mu$m, if interpreted as dust emission, would imply a very cold temperature, and it is possible that such excesses are due to contamination from background galaxies (\citealt{Krivov13_ColdDiscs}; \citealt{Gaspar14_ColdDiscs}). From \citet{Trilling07_BinaryDebris}, HD~46273 and HD~173608 were both deemed in that paper to have MIPS 70~$\mu$m excesses, but in both cases the photometry in that band, while a little high (2--3$\sigma$), is considerably below our excess threshold. Additionally, both were reobserved as part of the \texttt{OT2_gkennedy_2} survey, and the PACS 70~$\mu$m photometry is in line with the model for the stellar photosphere. Finally, while the \citet{Rodriguez12_BinaryDebris} systems were not included in Fig.~\ref{fig:sma_CDF_all}, it can be seen from Fig.~\ref{fig:rvsa} that none of these lie in the separation range of interest (and satisfy our excess criterion) either. There are two systems in our sample from that paper which fall in that separation range: HD~56986 and HD~73752. Both have slightly high (2--3$\sigma$) \textit{IRAS} 60~$\mu$m fluxes -- and consequently were previously thought to host discs -- but also MIPS 70~$\mu$m and PACS 100~$\mu$m fluxes that agree well with the stellar model. We are therefore satisfied that no excesses have been `missed' in the separation range in question, and that there is a genuine absence of debris discs there.

\begin{figure}
	\centering
    \hspace{-0.5cm}
	\includegraphics[width=0.5\textwidth]{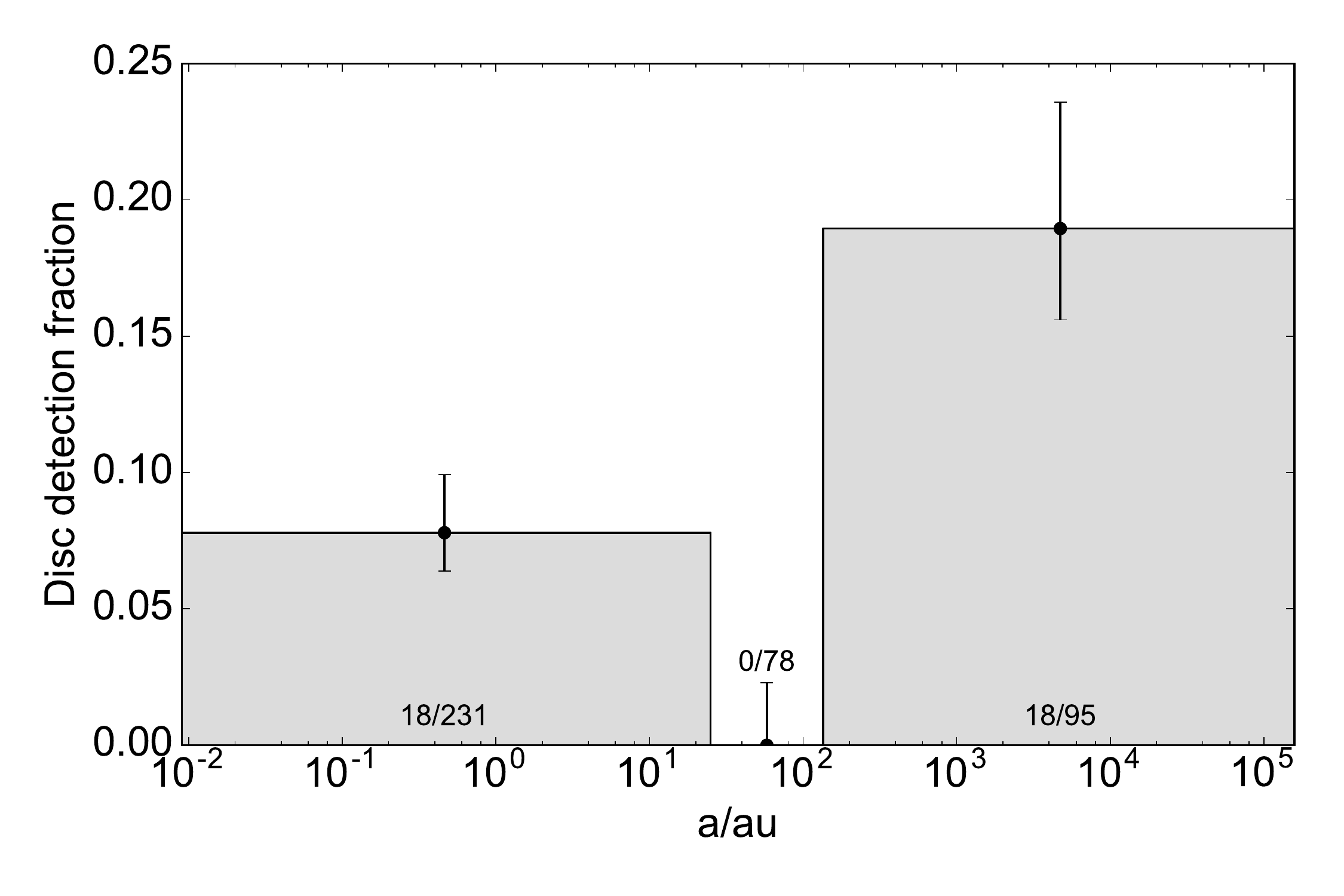}
	\caption{Fraction of unbiased systems in which we detect a debris disc, binned into three stellar separation ranges: $a<25~\mathrm{au}$, $25~\mathrm{au}<a<135~\mathrm{au}$ and $a>135~\mathrm{au}$. The annotations show the number of disc detections and stellar separations in each bin, with triple and higher order systems counted once per separation.}
	\label{fig:detfrac_sma}
\end{figure}

It is also instructive to examine the disc detection fraction as a function of stellar separation. This is illustrated in Fig.~\ref{fig:detfrac_sma}, where the separation bins were chosen as $a<25~\mathrm{au}$, $25~\mathrm{au}<a<135~\mathrm{au}$ and $a>135~\mathrm{au}$ to highlight the lack of discs in medium separation systems outlined above; the error bars show binomial uncertainties. This Figure shows that discs appear to be significantly more common in wide binaries, with a detection rate of $\frac{18}{95}=19^{+5}_{-3}\%$, than in those with close orbits, at $\frac{18}{231}=8^{+2}_{-1}\%$. Considering only the systems with well known orbits gives consistent results: 70 known orbit components lie in the closest semimajor axis bin; this includes all four with discs, giving a detection rate in this bin of $6^{+4}_{-2}\%$. The remaining 17 unbiased known orbit components lie in the medium separation bin. Our overall disc detection rate for all unbiased systems across all separations is $9^{+2}_{-1}\%$, consistent with the $11\pm3\%$ derived by \citet{Rodriguez15_BinaryDebris}. The detection rates per spectral type are $22^{+7}_{-5}\%$ for A stars, $11^{+3}_{-2}\%$ for F stars, $5^{+3}_{-1}\%$ for G stars and $7^{+4}_{-2}\%$ for K stars, with no discs detected around M stars; these results are again comparable with those of \citet{Rodriguez15_BinaryDebris}. 

\section{Discussion}
\label{sec:discussion}

In this section, we begin with a discussion of how our findings relate to the results of studies of protoplanetary discs and planets in binary systems. We then turn our attention to the issue of warm excesses in close binaries discussed by \citet{Matranga10_CloseBinaries}, and discuss what the \texttt{OT1_jdrake01_1} data add to our understanding of such systems.

\subsection{Relation to protoplanetary discs}
\label{sec:discussprotoplanetary}

Perhaps the most important result of this paper is the apparent lack of debris discs in systems with stellar separation $a$ between $\sim$25 and 135~au. This range of separations corresponds well to the typical disc radii that we derive, with the lower and upper quartiles of $r_{\mathrm{disc}}$ for our full sample being $\sim$30 and 140~au. Thus, the natural conclusion is that the lack of discs in medium separation binaries is a result of these systems clearing out their circumstellar or circumbinary material via dynamical perturbations. As detailed in section~\ref{sec:intro}, similar results have been hinted at in previous work (e.g. \citealt{Trilling07_BinaryDebris}), though not in a statistically significant way. 

The properties of protoplanetary discs in binary systems are also known to be dependent on stellar separation. It has been well established that the $\sim$mm fluxes of pre-main-sequence (PMS) binaries are significantly lower for systems with $1\lesssim a\lesssim100~\mathrm{au}$ than for $a\gtrsim100~\mathrm{au}$ (e.g. \citealt{Osterloh95_LowProtoFluxes}; \citealt{Jensen96_BinaryProtoDiscs}; \citealt{Andrews05_BinaryProtoDiscs}; \citealt{Duchene10_PltFormationBinaries}; \citealt{Kraus12_PPDMultiplicity}). This implies that systems with a separation comparable to typical disc sizes tend to have a lower protoplanetary disc mass, likely due to the dynamical influence of the binary companion. \citet{Cieza09_BinaryProtoDiscs} similarly found that PMS systems with $1\lesssim a\lesssim 40~\mathrm{au}$ are significantly less likely to have an excess at 8~$\mu$m (with the shorter wavelength probing the innermost part of a protoplanetary disc) than those with $40\lesssim a\lesssim 400~\mathrm{au}$. Since protoplanetary discs are the precursors of debris discs, in that the planetesimals comprising a debris disc are formed from material in the protoplanetary disc, we would expect to find a dearth of detectable debris discs for separations of order $\sim$1--100~au; this is indeed similar to our conclusion. 

If the properties of the debris discs we are studying are indeed influenced by stellar perturbations to their primordial protoplanetary discs, one might also expect circumbinary debris discs (for which truncation occurs from the inner edge) to have larger radii on average than circumstellar debris discs in binary systems (for which truncation occurs from the outer edge). We find that the median of $r_{\mathrm{disc}}$ is $75~\mathrm{au}$ for our $a<25~\mathrm{au}$ systems, compared with $55~\mathrm{au}$ for our $a>135~\mathrm{au}$ systems, consistent with this idea. This may be an interesting result, however given the relatively small number of discs it is not statistically significant, with a KS test on the disc radii of the close and wide binaries returning a $p$-value of 0.84.

We also found that the dust detection rate for the widest binaries in our sample, $19^{+5}_{-3}\%$, is comparable with single star rates (\citealt{Su06_AStarDiscs}; \citealt{Hillenbrand08_FEPS}; \citealt{Trilling08_DDSunLike}; \citealt{Eiroa13_DUNESDiscFrac}; \citealt{Sierchio14_MIPSPhot}; \citealt{Thureau14_HerschelA}; \citealt{Sibthorpe18_HerschelFGK}), while for the closest binaries debris discs are around half as common, with a detection rate of $8^{+2}_{-1}\%$. Statistical studies of discs around PMS binaries have not generally included systems with $a\lesssim 1~\mathrm{au}$ since PMS multiplicity surveys at such close separations are highly incomplete; we note, however, that at least several PMS binaries with $a\lesssim 1~\mathrm{au}$ are known to have high submillimetre fluxes (e.g. \citealt{Mathieu95_GWOri}, \citeyear{Mathieu97_DQTau}; \citealt{Kennedy19_HD98800}). Thus, it is not known whether protoplanetary disc properties in $a\lesssim 1~\mathrm{au}$ binaries are significantly different to those in $a\gtrsim100~\mathrm{au}$ binaries, and the reason for the lower debris disc incidence that we find in close binaries is not clear. It could be the case that close binaries tend to have lower protoplanetary disc masses than wide binaries, and thus a reduced ability to form planetesimals. Alternatively, it is possible that close binaries do form planetesimals as efficiently as wide binaries, but are able to drive faster collisional evolution of the resulting debris discs.

To summarise, medium separation binaries are lacking in circumstellar or circumbinary material from early in their lives. It appears that our conclusion that those on the main sequence lack detectable levels of dust is not predominantly a result of the dynamical evolution of planetesimals in the debris disc phase, but follows naturally from the lower masses of the primordial discs from which planetesimals form. Our result is ultimately due to the fact that the dynamical influence of the binary is important when $a$ is similar to typical values of $r_{\mathrm{disc}}$, but this influence leaves its observable signature on systems at an early stage.

\subsection{Relation to planets}
\label{sec:discussplanets}

The occurrence rate of planets in binaries is a strong function of stellar separation. Numerous studies have demonstrated that the separations of known planet-hosting binaries are significantly larger than those of field binaries, with the conclusion being that planet formation appears to be inhibited in systems closer than $\sim$10--100~au (e.g. \citealt{Bergfors13_PlanetsInBinaries}; \citealt{Wang14_PlanetsInBinaries}; \citealt{Kraus16_RuinousCloseBinaries}; \citealt{Fontanive19_PlanetsInBinaries}). Since the detection of a debris disc in a given system indicates that planetesimals were able to form therein, our result that discs are significantly less common for systems in the closest bin of Fig.~\ref{fig:detfrac_sma} than those in the widest bin may lead to the similar conclusion that planetesimal formation is hindered in systems with $a<25~\mathrm{au}$ (see also section~\ref{sec:discussprotoplanetary}).

Another potentially interesting result from the literature relates to even closer binaries: \citet{Welsh14_CircumbinaryKepler} noted that of the eight systems discovered by \textit{Kepler} to host circumbinary planets, none have binary periods shorter than $\sim$7~d, contrasting with the fact that the periods of most \textit{Kepler} eclipsing binaries are below 1~d. \citet{Fabrycky07_KozaiCycles} explored the idea that the closest binaries (those with periods less than $\sim$10~d, corresponding to semimajor axes below $\sim$0.1~au for Sun-like stars) are initially much wider, then have their orbits shrunk by Kozai cycles induced by a tertiary stellar companion combined with tidal friction. \citet{Martin15_NoCloseBinaryPlanets} suggested that this process could explain the apparent lack of circumbinary planets around very close binaries, since the third star would also perturb any circumbinary material, hindering planet formation and survival. 

From Fig.~\ref{fig:sma_CDF_all} it is clear that there is no `cutoff' separation below which no debris discs are found, in contrast with the circumbinary \textit{Kepler} planet results. In fact, the disc detection rate for the \texttt{OT1_jdrake01_1} sample of very close binaries is consistent with that for all systems with $a<25~\mathrm{au}$ (see section~\ref{sec:closebinaries}). This may suggest that the mechanism of Kozai cycles with tidal friction is not responsible for producing all (or the overwhelming majority) of the close binaries in our sample, as we would generally expect the planetesimals comprising debris discs to be cleared out in systems where this process has taken place. Thus, we expect that planets do indeed exist around some very close binaries and that such systems should be detected given a larger sample of circumbinary planets in the future.

\subsection{Infrared emission from close binaries}
\label{sec:closebinaries}

As noted earlier in this paper, \citet{Matranga10_CloseBinaries} studied a sample of ten close binaries and found warm infrared excesses in three of them using 24~$\mu$m photometry from MIPS and shorter wavelength photometry from \textit{Spitzer}'s Infrared Array Camera (IRAC; \citealt{Fazio04_IRAC}). They interpreted these warm excesses as possible emission from dust produced in collisions between planetary bodies. As such dust is not being continuously replenished, it should be relatively short-lived, and to detect it would require observing a system at a special point in its evolution. It would thus be somewhat surprising if 30\% of their sample were genuinely undergoing such an event at the time of observation. Our sample contains 50 systems with PACS data from the \texttt{OT1_jdrake01_1} survey, which was designed as a longer wavelength follow-up to \citet{Matranga10_CloseBinaries}, and in this subsection we consider whether these data offer any insight into the nature of the infrared emission from close binaries.

The \texttt{OT1_jdrake01_1} systems have a median separation of 0.05~au (or a median period of $\sim$4~d), and constitute a sample of the very closest binaries in our study. According to the criteria of section~\ref{sec:SED_modelling}, we detect formal 70~$\mu$m excesses in $\frac{4}{50}=8^{+6}_{-2}\%$ of these systems. While one of these excesses is the exceptionally warm HD~21242 (UX~Ari; see section~\ref{sec:unstablesys}), it is clear from Fig.~\ref{fig:fvsT} that very close binaries do not \textit{generally} have unusually warm dust temperatures; if this system is discounted then the detection rate is $6^{+5}_{-2}\%$. Regardless of whether HD~21242 is included, the \texttt{OT1_jdrake01_1} detection rate is in good agreement with the $8^{+2}_{-1}\%$ that we derived for the $a<25~\mathrm{au}$ systems in section~\ref{sec:statistics}, which suggests that even if there is some process in the closest binaries which tends to produce warm dust -- e.g. resonance sweeping caused by secular shrinkage of the binary orbit due to tidal interactions, as proposed by \citet{Matranga10_CloseBinaries} -- it does not significantly affect the levels of Kuiper belt-like cool dust in these systems. 

\begin{figure}
	\centering
    \hspace{-0.5cm}
	\includegraphics[width=0.5\textwidth]{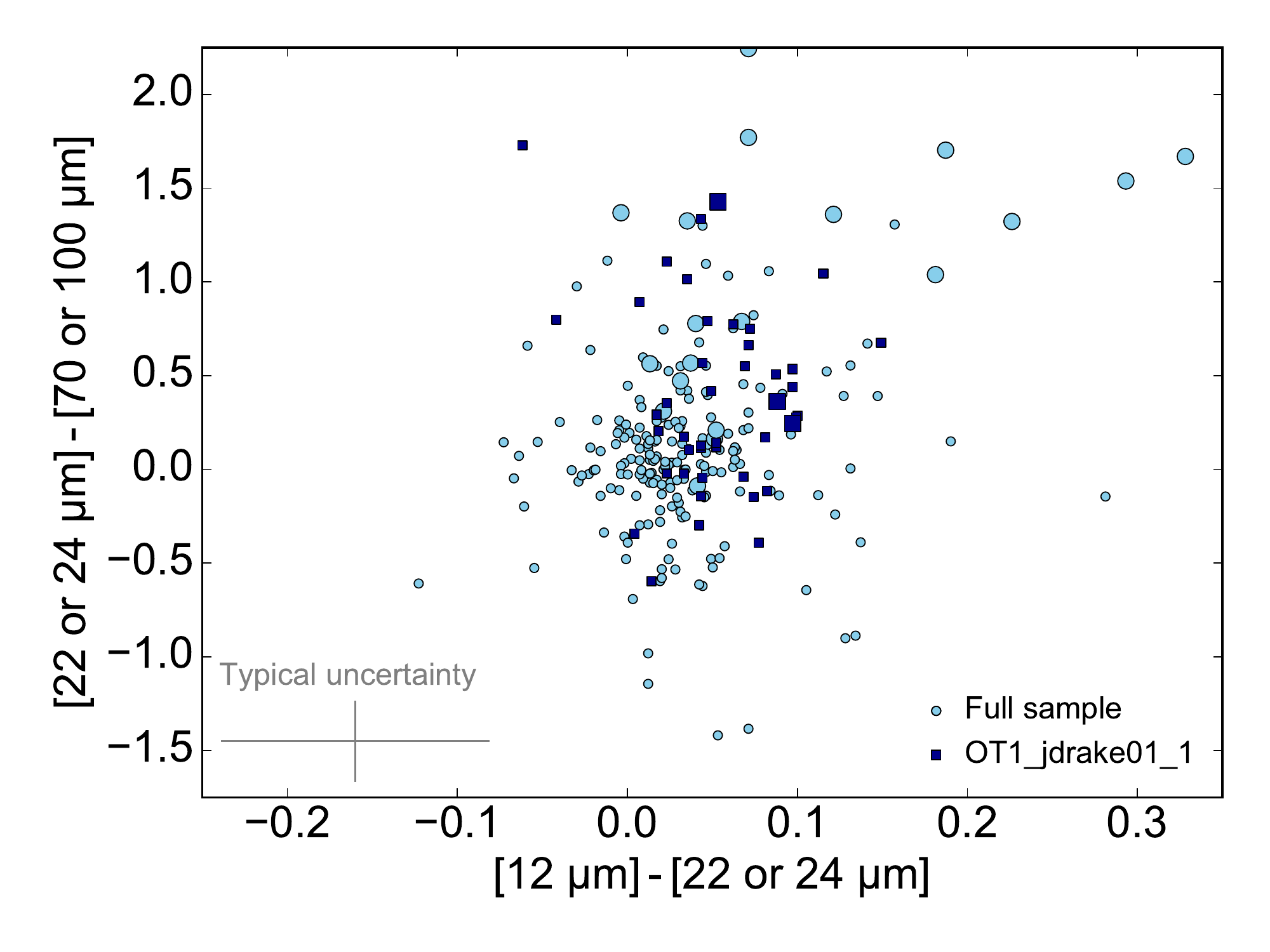}
	\caption{A colour-colour diagram for our full sample, with systems from the \texttt{OT1_jdrake01_1} survey highlighted. The horizontal axis shows WISE 12~$\mu$m magnitude minus WISE 22~$\mu$m magnitude where available and MIPS 24~$\mu$m otherwise. The vertical axis shows 22 or 24~$\mu$m magnitude minus far infrared magnitude, where for far infrared fluxes we preferentially used PACS 70~$\mu$m over MIPS 70~$\mu$m over PACS 100~$\mu$m. Systems with debris discs are plotted with larger symbols. The error bars show the median uncertainties in each colour. Systems towards the right (or the top) have SEDs which decrease less steeply than Rayleigh-Jeans between 12 and $\sim$22~$\mu$m (or $\sim$22 and $\sim$70~$\mu$m), while the SEDs of those towards the left (or the bottom) decrease \textit{more} steeply. The \texttt{OT1_jdrake01_1} sample appears to be systematically offset above and to the right of the rest of the systems.}
	\label{fig:col_col}
\end{figure}

Nonetheless, recall from section~\ref{sec:SED_modelling} that we do find a striking difference between the significance distributions of PACS 70~$\mu$m photometry for the \texttt{OT1_jdrake01_1} systems and the remainder of our sample, with the \texttt{OT1_jdrake01_1} fluxes appearing systematically higher than expected. To determine whether this offset is related to the apparent warm excesses found by \citet{Matranga10_CloseBinaries}, we can examine a colour-colour diagram, as shown in Fig.~\ref{fig:col_col}. In this Figure, the horizontal axis shows the magnitude in the WISE 12~$\mu$m band minus the $\sim$22~$\mu$m magnitude, where we used WISE 22~$\mu$m photometry where available and MIPS 24~$\mu$m otherwise. Similarly, the vertical axis shows $\sim$22~$\mu$m magnitude minus far infrared magnitude, for which we use PACS 70~$\mu$m fluxes where available, and MIPS 70~$\mu$m preferentially to PACS 100~$\mu$m where not. Each point represents a system in our sample, with the \texttt{OT1_jdrake01_1} systems highlighted as darker points. Systems which we deem to host a disc are plotted with larger symbols. The uncertainties in both colours, which depend on the fractional uncertainties in the fluxes, are found to have narrow distributions; typical error bars showing the median uncertainties are displayed in the Figure.

To interpret Fig.~\ref{fig:col_col} physically, note that systems with purely stellar SEDs have roughly equal magnitudes everywhere in their Rayleigh-Jeans tails and thus lie near the origin. Systems lying further to the right have fluxes that decrease less steeply than the Rayleigh-Jeans tail of a stellar SED between 12 and $\sim$22~$\mu$m; similarly for the fluxes of systems further towards the top between $\sim$22 and $\sim$70~$\mu$m. Conversely, systems below or to the left of the origin have \textit{more} steeply declining fluxes over the relevant wavelength range. Not all systems are visible in the plot; the brightest excesses lie outside the range shown.

In Fig.~\ref{fig:col_col} it can be seen that as a population, the \texttt{OT1_jdrake01_1} systems are concentrated above and to the right of the rest of the sample. That is, these systems tend to have SEDs which are declining less steeply than for a pure Rayleigh-Jeans tail both from 12 to $\sim$22~$\mu$m and from $\sim$22 to $\sim$70~$\mu$m. This situation can be contrasted with the case of HD~69830, a single star with an excess at $\sim$22~$\mu$m (\citealt{Beichman05_HD69830}) but \textit{not} at $\sim$70~$\mu$m (\citealt{Marshall14_PlanetsAndDebris}). The excess of HD~69830 has been interpreted as due to debris from a destructive collision of a $\sim$30~km--sized asteroid (\citealt{Beichman11_HD69830}), similar to the scenario proposed by \citet{Matranga10_CloseBinaries} to explain the infrared emission from close binaries. Its excess peaks at around 20~$\mu$m, with its SED falling less steeply than Rayleigh-Jeans from 12 to $\sim$22~$\mu$m, then \textit{more} steeply from $\sim$22~$\mu$m to $\sim$70~$\mu$m, and would thus lie to the right and \textit{below} the bulk population in Fig.~\ref{fig:col_col}. We would expect the SEDs of other systems experiencing such a collision to have similar shapes to that of HD~69830, peaking at a few tens of $\mu$m and then declining more steeply than a black body. This is because a collision between two massive bodies would be expected to produce an excess of small grains compared with a steady-state debris disc, meaning that in such cases $\lambda_0$, the wavelength beyond which the SED decreases more steeply (see section~\ref{sec:SED_modelling}), should be relatively small. In addition, the presence of silicate emission features at $\sim$10 and 20~$\mu$m (see e.g. \citealt{Matthews14_Review}) could act to steepen the SED between mid and far infrared wavelengths. We note that of the \texttt{OT1_jdrake01_1} sample, only HD~8997, HD~118216 and HIP~3362 have \textit{Spitzer} IRS spectra, but none of these show the silicate features which we might expect to see if an abundance of small grains were present.

In fact, the excesses proposed by \citet{Matranga10_CloseBinaries}, being exceptionally hot (1000 -- 2000~K), peak at even shorter wavelengths than for HD~69830 ($\sim$5$\mu$m). We might expect the SEDs of systems with such hot excesses to be falling more steeply than Rayleigh-Jeans even from 12 to $\sim$22~$\mu$m, placing them toward the left side of Fig.~\ref{fig:col_col}. In either case, while it appears from Fig.~\ref{fig:col_col} (as well as Figs~\ref{fig:chihist_nodrake} and \ref{fig:chihist_drake}) that the SEDs of the \texttt{OT1_jdrake01_1} sample are indeed somehow different from our bulk sample, their positions in this Figure do not seem to support the idea that massive collisions are much more common for close binaries than for wider binaries or single stars. 

The above discussion is based on the assumption that the dust is optically thin, and that its emission is therefore well described by the modified black body function defined in section~\ref{sec:SED_modelling}. The \texttt{OT1_jdrake01_1} SEDs may in fact be consistent with optically thick dust, as the emission of such dust is better described by an unmodified black body. However, the short collisional lifetime of warm dust would still be an issue, perhaps even more so if the dust were required to be present at high enough density to be optically thick.

To summarise, the expected transience of warm dust from giant collision events, combined with the shapes of the infrared spectra of the \texttt{OT1_jdrake01_1} systems as determined from a colour-colour diagram, indicate that further work is required to understand the true nature of the infrared emission from close binaries. It appears that such binaries are subject to some unknown systematic effect causing larger than expected fluxes in the Rayleigh-Jeans tails of their SEDs, which could cause warm dust to appear more common than it really is.

\section{Conclusions}
\label{sec:conclusions}

We have assembled a sample of 341 binary and multiple star systems, and modelled their SEDs to identify and characterise debris discs in these systems. Our sample, defined in section~\ref{sec:sample}, is a combination of the systems studied by \citet{Trilling07_BinaryDebris}, \citet{Rodriguez12_BinaryDebris} and \citet{Rodriguez15_BinaryDebris}, and the systems targeted by the \textit{Herschel} programmes \texttt{OT2_gkennedy_2} (a survey of visual binaries with well known orbits) and \texttt{OT1_jdrake01_1} (a survey of close binaries).

To each SED we fitted a model with both a stellar component and a modified black body dust component. As outlined in section~\ref{sec:SED_modelling}, we defined an infrared excess criterion based on the measured distributions of photometric significances, such that systematic effects and underestimated or overestimated uncertainties are accounted for. Systems with significant excesses in the MIPS 70~$\mu$m, PACS 70~$\mu$m and/or PACS 100~$\mu$m bands were considered to host a debris disc. In section~\ref{sec:discproperties} we found that there are 38 such systems. We converted the dust temperatures from SED fitting into black body radii $r_{\mathrm{bb}}$, then corrected these for the effect of the inefficient emission of small grains using the prescription of \citet{Pawellek15_TrueRadius}. The resulting radii $r_{\mathrm{disc}}$ were used in section~\ref{sec:stability} to assess the dynamical stability of the discs according to the criteria of \citet{Holman99_Stability}. Nine of the discs appear based on a simple analysis of their radii and host binary separations to be unstable, and we discussed each of these systems in detail in section~\ref{sec:unstablesys}. Considering the expected short lifetime of unstable dust, we are hesitant to conclude that the apparent instability is real. Uncertainty in the true disc radius due to scatter in the correction factor from $r_{\mathrm{bb}}$ to $r_{\mathrm{disc}}$, uncertainty in the boundary of the unstable region (which depends on the binary eccentricity and mass ratio), and projection of the binary semimajor axis could be responsible for stable systems appearing to be unstable.


We found in section~\ref{sec:statistics} that the separations of the disc-bearing and disc-free systems are drawn from different distributions at a confidence level of $99.4$\%, higher than the $91$\% found previously by \citet{Rodriguez15_BinaryDebris}. We thus concluded that the binary separation $a$ influences the presence of detectable levels of debris in a statistically significant way, in contrast with the conclusion of \citet{Rodriguez15_BinaryDebris}. The disc detection rate is $19^{+5}_{-3}\%$ for systems with $a>135~\mathrm{au}$, comparable with published results for single stars. Only $8^{+2}_{-1}\%$ of those with $a<25~\mathrm{au}$ have a detectable disc, which suggests that either planetesimal formation is inhibited in binaries closer than a few tens of au, or that such systems are able to drive faster collisional evolution in their debris discs than wide binaries. We did not find any significant infrared excesses in systems with $25~\mathrm{au}<a<135~\mathrm{au}$. Our main conclusion is thus that there is a lack of debris discs in medium separation systems.

We compared our results with studies of pre-main-sequence binaries in section~\ref{sec:discussprotoplanetary}, concluding that the lack of debris discs in medium separation systems is likely a result of their precursors, protoplanetary discs, being dynamically cleared of material when the stellar separation is similar to typical disc radii. In section~\ref{sec:discussplanets} we discussed how our conclusion that binaries with $a<25~\mathrm{au}$ are less likely to host a detectable debris disc relates to the results of studies of known planet hosting binaries, which have suggested that planet formation is hindered in binaries closer than a few tens of au. We then considered the non-detection by \textit{Kepler} of circumbinary planets in very short period binaries, and suggested that our detection of discs in such systems implies that future observations should in fact find such planets. Finally, in section~\ref{sec:closebinaries} we explored the idea of frequent planetary collisions in close binaries proposed by \citet{Matranga10_CloseBinaries}, concluding that the closest binaries in our sample do not tend to have the colours we would expect in such a scenario.

\section*{Acknowledgements}

We thank the reviewer for a report that helped to improve the quality of the paper. BY acknowledges the support of an STFC studentship. GMK is supported by the Royal Society as a Royal Society University Research Fellow.




\bibliographystyle{mnras}\input{refs.bbl}



\appendix
\section{Resolved image modelling of HD~95698}
\label{app:imagemodel}

A single disc -- that of HD~95698 -- was resolved in the \texttt{OT2_gkennedy_2} observations. Here we outline the method we used to model this disc, and present the results of this modelling.

\begin{figure*}
	\centering
	\includegraphics[width=0.95\textwidth]{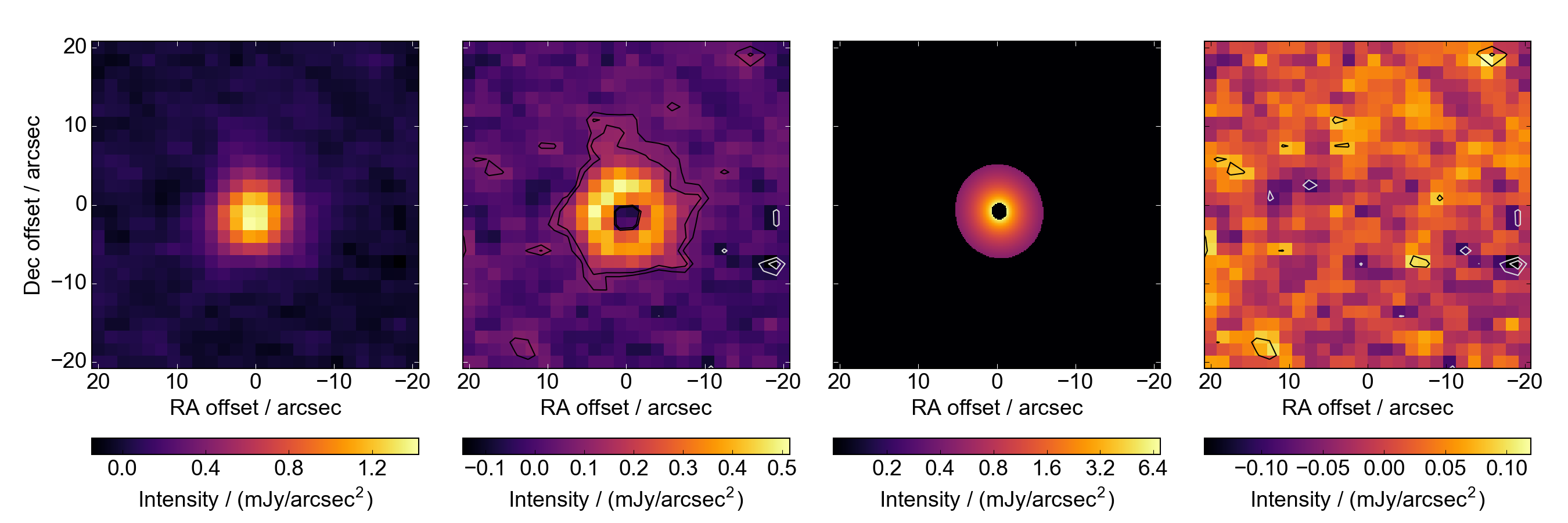}
	\caption{Far left: cutout of the PACS 70~$\mu$m image of HD~95698 from the \texttt{OT2_gkennedy_2} survey. Centre left: residuals following a PSF subtraction, with the PSF normalised to the brightest pixel in the observed image. The contours show $\pm$2 and 3 times the root mean square intensity. The ring of significant emission which remains indicates that the disc is resolved. Centre right: unconvolved high resolution image of the minimum $\chi^2$ model as determined using the Markov Chain Monte Carlo method. For clarity, we do not plot the central pixel containing the stellar and unresolved fluxes, and the colour scale is logarithmic with the background level arbitrarily set to $0.1~\mathrm{mJy~arcsec}^{-2}$. Far right: residuals of the minimum $\chi^2$ model, with overplotted contours of $\pm$2 and 3 times the root mean square intensity. There do not appear to be any significant structures in the residuals, indicating that the model explains the data well. }
	\label{fig:imagemodel}
\end{figure*}

A cutout of the 70~$\mu$m image of HD~95698 from that survey is shown in the leftmost panel of Fig.~\ref{fig:imagemodel}. The centre left panel of that Figure shows the residuals following a PSF subtraction. We use an observation of the \textit{Herschel} calibration star HD~164058 (gamma Dra), normalised and rotated to the appropriate orientation, as a PSF. For the subtraction, we scaled the PSF such that its peak intensity is equal to the peak intensity in the image. We allowed the PSF to have a variable offset since the \textit{Herschel} pointing is not perfect, and the residuals shown correspond to the offset which gives the best fit. A prominent ring of emission remains, indicating that the disc in this system is indeed resolved.

One pixel in the image covers 1.6$^{\prime\prime}$, and given that the system is at a distance of 57.4~pc (\citealt{vanLeeuwen07_Hipparcos}) this corresponds to 92~au per pixel. Given the low spatial resolution of the data, we generate our model images on a higher resolution grid with $13\times13$ smaller pixels per PACS pixel. Our model for the resolved disc is axisymmetric and optically thin, and assumes that the surface density is proportional to $r^{-1}$, where $r$ is the distance from the centre of mass of the binary, and thus that the surface brightness is proportional to $r^{-1.5}$ (assuming that the dust temperature at a distance $r$ is proportional to $r^{-0.5}$, as is the case for black body grains). We also tried an alternative model in which the surface brightness varies as $r^{-\alpha}$, with $\alpha$ as a free parameter. However, we found that $\alpha$ was poorly constrained, with the model preferring to have unphysically steep profiles and unphysically large outer disc radii such that the disc emission drops below the noise floor. Thus, we ultimately decided to fix $\alpha$ at a value of 1.5, which we consider to be physically reasonable.

The model has as free parameters the total flux of the resolved disc $F_{\mathrm{res}}$, inner and outer edges $r_1$ and $r_2$, inclination $I$, and position angle of the major axis $\theta$ (measured anticlockwise from north). Our SED fitting gives the expected stellar flux as $5.372~\mathrm{mJy}$. It is known that for PACS data the flux obtained from PSF fitting will underestimate the true value as some flux is lost in the wings of the PSF during image processing; at 70~$\mu$m this underestimation is by a factor of around 1.16 (\citealt{Kennedy12_99Her}). Thus, at the pixel in the centre of the disc we set the flux to $F_{\mathrm{star}}+F_{\mathrm{unres}}$, where $F_{\mathrm{star}}=\frac{5.372}{1.16}~\mathrm{mJy}$, and $F_{\mathrm{unres}}$ is a free parameter allowing for an unresolved component of the excess. Including an arbitrary two-dimensional offset -- which is unrelated to the properties of the disc but accounts for non-perfect \textit{Herschel} pointing -- there are therefore eight free parameters in total.

After resampling the PSF to the higher-resolution model grid, we convolve the model with the PSF; the resulting synthetic image is then resampled back to the PACS pixel size for comparison with the observations. We calculate the goodness of fit $\chi^2$ -- which is unrelated to the significance $\chi$ defined in equation~(\ref{eqn:chidef}) -- as follows:

\begin{equation}\label{eqn:chisqdef}
    \chi^2 = \sum_{i}{\frac{(F_{\mathrm{obs},i}-F_{\mathrm{model},i})^2}{\sigma^2}}.
\end{equation}

Here $F_{\mathrm{obs},i}$ and $F_{\mathrm{model},i}$ are the observed and model fluxes in pixel $i$, and $\sigma$ is the corresponding uncertainty. We calculate the sum over a $26\times26$ pixel box centred on the star. To estimate the appropriate value of $\sigma$, we first took the root mean square (RMS) of the flux per pixel in an empty $20\times20$ pixel patch of sky near the star, giving $0.097~\mathrm{mJy}$. As outlined in \citet{Kennedy12_99Her}, the noise in a PACS image is correlated, and this effect can be accounted for by scaling the RMS by the factor defined in equation~(9) of \citet{Fruchter02_Drizzle}. The ratio of natural to actual pixel scales in the image we are modelling is $\frac{3.2^{\prime\prime}}{1.6^{\prime\prime}}=2$, which leads to the result $\sigma=0.23~\mathrm{mJy}$. To find the best fitting model, we first estimated the parameters which minimise $\chi^2$ using differential evolution, an algorithm designed to find the global minimum of a function, avoiding the potential issue of instead returning a local minimum. To quantify the posterior distributions of the model parameters, we then ran the Markov Chain Monte Carlo (MCMC) algorithm implemented by \citet{ForemanMackey13_MCMC}, with a likelihood function $e^{-\chi^2/2}$. We used 250 walkers initialised close to the previously identified minimum $\chi^2$ solution, and ran the algorithm for 1000 steps. Inspection of the individual chains led us to discard the first 700 steps as the burn-in period. 

The resulting parameters can by quantified by the median values of the posterior distributions, with uncertainties corresponding to the 16th and 84th percentiles: $F_{\mathrm{unres}}=19^{+14}_{-13}~\mathrm{mJy}$, $F_{\mathrm{res}}=99^{+14}_{-15}~\mathrm{mJy}$, $r_1=101^{+72}_{-38}~\mathrm{au}$, $r_2=318^{+43}_{-54}~\mathrm{au}$, $I=16^{+11}_{-11}~\mathrm{deg}$ and $\theta=5^{+38}_{-46}~\mathrm{deg}$. As expected, $F_{\mathrm{unres}}$ and $F_{\mathrm{res}}$ are anticorrelated, with their sum being well constrained to $118^{+3}_{-3}~\mathrm{mJy}$. Similarly, $r_1$ is correlated with $F_{\mathrm{unres}}$; for a greater amount of unresolved flux, the resolved disc needs to contribute less flux close to the star. Including the stellar flux, and applying the correction factor of 1.16 for lost flux as outlined above, our modelling implies a total flux of $144\pm3~\mathrm{mJy}$, which is just consistent with the $154\pm7~\mathrm{mJy}$ we measured using aperture photometry for the purposes of SED fitting. We show the unconvolved image corresponding to the best-fitting (i.e. minimum $\chi^2$) model in the centre right panel of Fig.~\ref{fig:imagemodel}, and the resulting residuals in the rightmost panel; the residuals show that this model explains the data well. Its parameters are $F_{\mathrm{unres}}=3~\mathrm{mJy}$, $F_{\mathrm{res}}=115~\mathrm{mJy}$, $r_1=60~\mathrm{au}$, $r_2=345~\mathrm{au}$, $I=22~\mathrm{deg}$ and $\theta=15~\mathrm{deg}$. The best fitting values of $F_{\mathrm{unres}}$, $F_{\mathrm{res}}$ and $r_1$ are notably different from their medians, which is a result of the correlations between the parameters discussed above. While the stellar orbit of HD~95698 is not well known and we are thus unable to test for misalignment between the disc and the binary plane, the radii derived from image modelling are useful for comparison with the disc size derived from SED fitting (see section~\ref{sec:unstablesys}).



\bsp	
\label{lastpage}
\end{document}